\documentclass[aps,prb,twocolumn,longbibliography,superscriptaddress]{revtex4-2}
\usepackage{graphicx}
\usepackage{bm}
\usepackage{wrapfig}
\usepackage{color}
\usepackage{natbib}
\usepackage{hyperref}
\usepackage{amssymb}
\usepackage{amsmath}
\usepackage{geometry}
\usepackage{comment}

\begin{document}
%%%%%%%%%%%%%%%%%%%%%%%%%%%%%%%%%%%%%%

%%%%%%%%%%%%%%%%%%%%%%%%%%%%%%%%%%%%%%
\title{Optical control of topological memory based on orbital magnetization}
%%%%%%%%%%%%%%%%%%%%%%%%%%%%%%%%%%%%%%

\author{Sergey S. Pershoguba}

\affiliation{Department of Physics and Astronomy, University of New Hampshire, Durham, New Hampshire 03824, USA}

\author{Victor M. Yakovenko}

\affiliation{JQI, Department of Physics, University of Maryland, College Park, Maryland 20742, USA}

% \date{PRL submission 2021-6-19; v2 arXiv 2021-6-19; v3 arXiv 2021-6-28; v4 PRB submission 2022-1-14}

\date{February 18, 2022}

%%%%%%%%%%%%%%%%%%%%%%%%%%%%%%%%%%%%%%%%%%%%%%%%%%%%%%%%%%%%%%
\begin{abstract} \noindent
Under suitable conditions, some twisted graphene multilayers and transition-metal dichalcogenides become Chern insulators, exhibiting the anomalous quantum Hall effect and orbital magnetization due to spontaneous valley polarization. We study the interaction of a Chern insulator with circularly polarized light.  The interaction energy contains an antisymmetric term that couples to the helicity of incident light.  For a two-band Chern insulator, this term is expressed as an integral involving the Berry curvature of the system.  Taking advantage of this interaction, we propose an experimental protocol for switching topological memory based on orbital magnetization by circularly polarized light.  Moreover, two laser beams of opposite circular polarization can nucleate domains of opposite magnetization and thus produce an optically configurable domain wall carrying topologically protected chiral edge modes.
\end{abstract}
%%%%%%%%%%%%%%%%%%%%%%%%%%%%%%%%%%%%%%%%%%%%%%%%%%%%%%%%%%%%%%

\maketitle

%%%%%%%%%%%%%%%%%%%%%%%%%%%%%%%%%%%%%%%%%%%%%%%%%%%%%%%%%%%%%%%%%%%%%%%%%%%%%
\section{Introduction}
%%%%%%%%%%%%%%%%%%%%%%%%%%%%%%%%%%%%%%%%%%%%%%%%%%%%%%%%%%%%%%%%%%%%%%%%%%%%%

There has been much interest lately in heterostructures of twisted graphene multilayers \cite{Andrei2021}. At a small interlayer twist angle $\sim 1^\circ$, they form a periodic moir{\'e} superlattice with a large unit cell size $a_m \sim 10$~nm and exhibit flat electronic energy bands~\cite{Bistritzer2011}.  Therefore these systems provide a versatile and highly tunable platform for studying electron correlations.  Among many interesting phases, Chern insulators, which have a nonzero integer topological Chern number $C$ and exhibit quantized Hall resistance $R_H$, were recently observed \cite{Lu2019,Sharpe2019,Chen2020,Serlin2020,Polshyn2020}.  They originate from spontaneous valley polarization induced by electron correlations, resulting in orbital magnetization $\bm M$ due to the incipient Berry curvature of graphene \cite{Song2015}.  Experiments \cite{Lu2019,Sharpe2019,Chen2020,Serlin2020,Polshyn2020} have demonstrated that the sign of $\bm M$ can be switched by weak pulses of electric or magnetic fields, or electric currents, thus realizing nonvolatile topological memory \cite{Polshyn2020}.

\begin{figure} 
\includegraphics[width=0.8\linewidth]{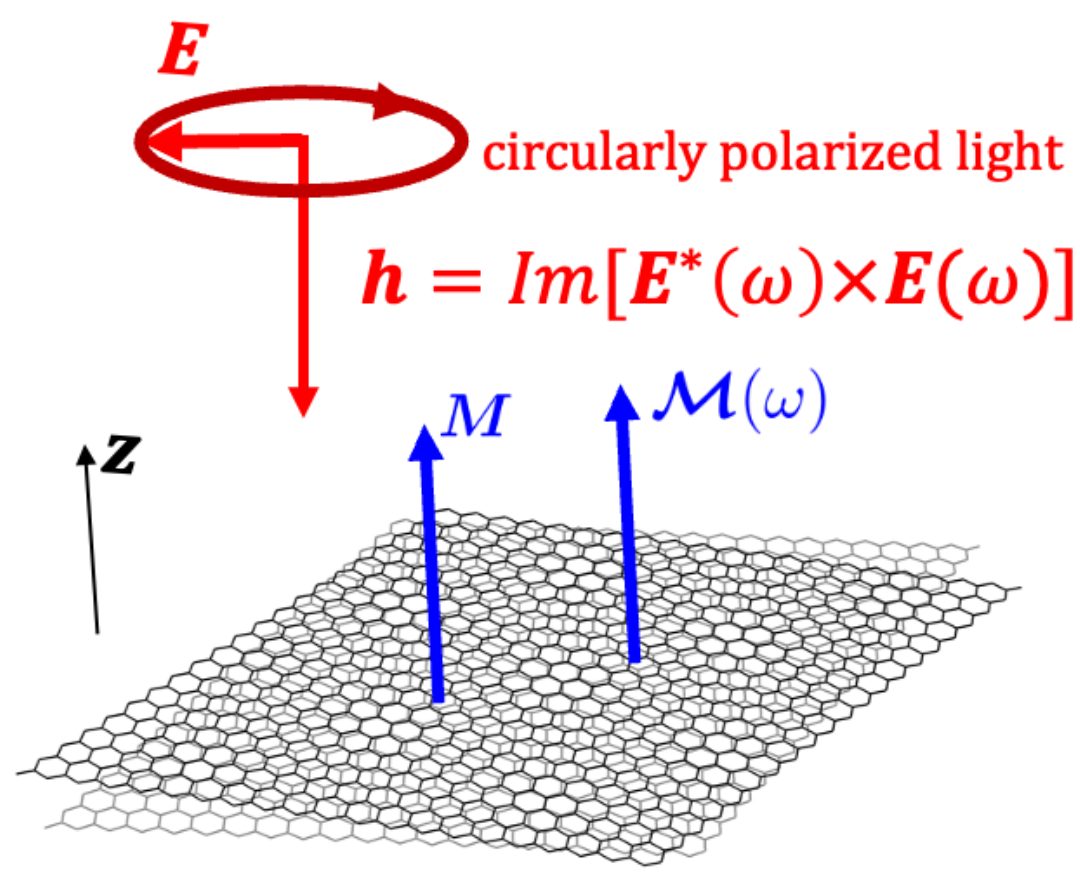}
\caption{Experimental setup for controlling orbital magnetization $\bm M$ of a Chern insulator by circularly polarized light of helicity $\bm h$.  The axial vector $\bm{\mathcal M}(\omega)$ characterizes the antisymmetric part of the dynamical polarizability tensor (\ref{chi-M}).}
\label{fig:setup}
\end{figure}

Here we propose all-optical control of the orbital topological memory. Consider circularly polarized light normally incident onto the surface of a Chern insulator, as shown in  Fig.~\ref{fig:setup}. Its helicity $\bm h$, parallel to the direction of light propagation, acts as an effective magnetic field.  It couples to orbital magnetization $\bm M$ of the Chern insulator, which is perpendicular to the layers.  The sign of the interaction energy is determined by $\bm h\cdot\bm M$, thus helicity of light makes one sign of $\bm M$ more energetically favorable, which can switch the memory.

%%%%%%%%%%%%%%%%%%%%%%%%%%%%%%%%%%%%%%%%%%%%%%%%%%%%%%%%%%%%%%%%%%%%%%%%%%%%%
%%%%%%%%%%%%%%%%%%%%%%%%%%%%%%%%%%%%%%%%%%%%%%%%%%%%%%%%%%%%%%%%%%%%%%%%%%%%

Magnetization switching by light was observed in spin-based magnets both in helicity-dependent and helicity-independent manner: see reviews by St\'ephane Mangin, Ch.~9 ``Ultrafast magnetisation dynamics'' in Ref.~\cite{Roadmap2017} and by Andrei Kirilyuk, Ch.~5 ``All-optical magnetization reversal'' in Ref.~\cite{Vedmedenko2020}.  In contrast, we focus on Chern insulators, which have the advantage of direct electric-dipole coupling to light due to their orbital nature.  The interaction between orbital magnetization and circularly polarized light is general and not limited to twisted graphene multilayers.  There have been extensive studies of valley-selective optics in transition-metal dichalcogenides \cite{Mak2018}.  The ac Stark shift of the exciton energy was observed in WSe$_2$ \cite{Kim2014} and WS$_2$ \cite{Sie2015}, as well as the Bloch-Siegert shift in WS$_2$ \cite{Sie2017}.  Recently, spontaneous valley polarization was observed in the AB-stacked MoTe$_2$/WSe$_2$ bilayers \cite{Mak2021}, reminiscent of the similar phenomenon in twisted graphene multilayers.

The paper is organized as follows.  We start with a phenomenological symmetry analysis of the ac Stark effect in Sec.~\ref{sec:symmetry}, followed by a description of the proposed experimental protocol in Sec.~\ref{sec:experiment}.  These two Sections involve minimal mathematics and focus on qualitative arguments.  They are particularly recommended for experimentalists and general readers primarily interested in a physical picture, rather than mathematical details.  The subsequent Sections present a microscopic mathematical derivation of the ac Stark energy shift, first in Sec.~\ref{sec:discrete} for a localized system with a discrete energy spectrum, such as an atom, and then for a crystalline Chern insulator in Sec.~\ref{sec:Chern}.  A connection between the dynamical polarizability and the ac Hall conductivity is established in Sec.~\ref{sec:acHall}.  A contribution to the Stark energy shift from the edge states in a Chern insulator is considered in Sec.~\ref{sec:edge}.  Applications of the general theoretical results to actual materials are discussed in Sec.~\ref{sec:valley} for valley-selective optics on the honeycomb lattice and to twisted graphene multilayers in Sec.~\ref{sec:graphene}, ending by Conclusions in Sec.~\ref{sec:conclusions}.

%%%%%%%%%%%%%%%%%%%%%%%%%%%%%%%%%%%%%%%%%%%%%%%%%%%%%%%%%%%%%%%%%%%%%%%%%%%%%
\section{Phenomenological symmetry analysis}
\label{sec:symmetry}
%%%%%%%%%%%%%%%%%%%%%%%%%%%%%%%%%%%%%%%%%%%%%%%%%%%%%%%%%%%%%%%%%%%%%%%%%%%%%

Suppose a monochromatic ac electric field $\bm E(t)$ is applied to the system
\begin{equation}  \label{E(t)_app}
  \bm E(t) = \frac12 \, \left[\bm E(\omega)\,e^{-i\omega t} 
  + \bm E(-\omega)\,e^{i\omega t}\right],
\end{equation}
where $\bm E(-\omega)=\bm E^\ast(\omega)$.  The electric field couples to the electric dipole moment in the Hamiltonian of the system
\begin{equation}  \label{E.r_app}
  H = H_0 - e \bm r\cdot\bm E(t),
\end{equation}
where $H_0$ is the bare Hamiltonian, and the operator $\bm r$ is either the coordinate of a single electron or the sum of coordinates for many electrons.  As a result, a nonzero expectation value of the electron dipole operator is induced
\begin{equation}  \label{d(t)_app}
  \bm d(t) = e\langle\bm r\rangle   
  = \frac{\bm d(\omega)\,e^{-i\omega t} + \bm d(-\omega)\,e^{i\omega t}}{2}.
\end{equation}
It is linearly related to the electric field by the dynamical polarizability tensor $\chi^{\alpha\beta}(\omega)$:
  \begin{equation}  \label{d(omega)_app}
  d^\alpha(\omega) = \chi^{\alpha\beta}(\omega) \, E_\beta(\omega),
\end{equation}
where $\alpha$ and $\beta$ are the Cartesian indices in three dimensions (3D).  The ac Stark shift \cite{Delone1999,haas2006,Kobe1983} is the corresponding change $U$ of the energy of the system
\begin{align}  \label{energy_shift_def}
  U = - \frac14 {\rm Re}\left[\chi^{\alpha\beta}(\omega)E_{\alpha}(-\omega)E_{\beta}(\omega) \right] \equiv U_s + U_a.
\end{align}
The polarizability tensor $\chi^{\alpha\beta}(\omega)$ can be decomposed into the symmetric and antisymmetric parts
\begin{align}
  &\chi^{\alpha\beta}_s(\omega) = \frac12 \left[\chi^{\alpha\beta}(\omega) 
  + \chi^{\beta\alpha}(\omega)\right],
  \label{chi_s} \\
  & \chi^{\alpha\beta}_a(\omega) = \frac12 \left[\chi^{\alpha\beta}(\omega) 
  - \chi^{\beta\alpha}(\omega)\right],
  \label{chi_a}
\end{align}
thus generating the corresponding contributions $U_s$ and $U_a$ to the energy shift (\ref{energy_shift_def})
\begin{align}
  &U_s = -\frac14 {\rm Re}\left[E_\alpha^\ast(\omega)E_\beta(\omega)\right]
  {\rm Re}\left[\chi^{\alpha\beta}_s(\omega)\right],
  \label{energy_shift_sym} \\
  & U_a =\frac14 {\rm Im}\left[E_\alpha^\ast(\omega)E_\beta(\omega)\right]
  {\rm Im}\left[\chi^{\alpha\beta}_a(\omega)\right]. 
  \label{energy_shift_asym}
\end{align}
 The symmetric contribution $U_s$ represents the conventional ac Stark energy shift~\cite{Kobe1983,Delone1999,haas2006} of no interest to us. The antisymmetric contribution $U_a$ is less common and is permitted only for systems with broken time-reversal symmetry. The symmetric tensor ${\rm Re}\,\chi_s^{\alpha\beta}(\omega)={\rm Re}\,\chi_s^{\alpha\beta}(-\omega)$ is even, whereas the antisymmetric tensor ${\rm Im}\,\chi_a^{\alpha\beta}(\omega)=-{\rm Im}\,\chi_a^{\alpha\beta}(-\omega)$ is odd upon time reversal \cite{LandauVol5}.  Pitaevskii~\cite{Pitaevskii1961} noted that the antisymmetric part $\chi_a^{\alpha\beta}$ is permitted in the presence of an external magnetic field.  The subject advanced further with the experimental discovery of the inverse Faraday effect \cite{vanderZiel1965}, where circularly polarized light induces magnetization in a material.  A theory of the light-induced spin magnetization of Eu$^{++}$ ions in a solid-state matrix was developed in Ref.~\cite{Pershan1966}.
 
The recently discovered orbital Chern insulators \cite{Sharpe2019,Lu2019,Serlin2020,Chen2020,Polshyn2020} break time-reversal symmetry spontaneously and, thus, should have a nonzero $\chi_a^{\alpha\beta}$, which is the subject of our paper.  It is convenient to represent the antisymmetric tensor $\chi_a^{\alpha\beta}(\omega)$ via a dual vector $\bm {\mathcal M}(\omega)$ 
 \begin{equation}  \label{chi-M}
   \chi_a^{\alpha\beta}(\omega) = 
   - i\,\epsilon^{\alpha\beta\gamma}\,{\mathcal M}_\gamma(\omega),
 \end{equation}
 where $\epsilon^{\alpha\beta\gamma}$ is the totally antisymmetric tensor in 3D.  Then the corresponding energy contribution is
\begin{align}  \label{energy_shift}
    U_a =  -\frac14 {\rm Im}\left[\bm E^\ast(\omega)\times \bm E(\omega)\right]
    \cdot {\rm Re}\,\bm{\mathcal M}(\omega).
\end{align}
Since ${\rm Re}\,\bm {\mathcal M}(\omega)$ is an odd function of frequency $\omega$, one may expect to write it at low $\omega$ as ${\rm Re}\,\bm {\mathcal M}(\omega) \approx \omega \bm {\mathcal M}'(0)$.  Then, in time domain for slowly changing $\bm E(t)$, Eq.~(\ref{energy_shift}) would be
\begin{align}  \label{energy_shift_time_domain}
    U_a = \frac12 \, \langle\partial_t\bm E(t)\times \bm E(t)\rangle_t 
    \cdot \bm{\mathcal M}'(0), 
\end{align}
where $\langle\ldots\rangle_t$ denotes time averaging.  Equations (\ref{energy_shift}) and (\ref{energy_shift_time_domain}) show that the sign of the energy $U_a$ is determined by the scalar product of the vector $\bm {\mathcal M}(\omega)$ characterizing the system and the helicity 
\begin{align}  \label{helicity}
  \bm h(\omega) = {\rm Im}\left[\bm E^\ast(\omega)\times \bm E(\omega)\right]
\end{align}
of circularly polarized electric field, which is parallel to the vector $\bm E(t) \times \partial_t\bm E(t)$.  The vector $\bm {\mathcal M}(\omega)$ is a time-reversal-odd axial vector, so it has the same symmetry as the static magnetization $\bm M$ of the system.  Both $\bm{\mathcal M}(\omega)$ and $\bm M$ originate from the Berry curvature (as discussed later in the paper), but they are not equivalent.  

Combining Eqs.~(\ref{chi-M}) and Eq.~(\ref{d(omega)_app}), we find a dipole moment that is perpendicular to both the electric field $\bm E(\omega)$ and the vector $\bm{\mathcal M}(\omega)$, like in the Hall effect,
\begin{equation} \label{dEM_app}
  \bm d(\omega) = - i \bm E(\omega) \times \bm{\mathcal M}(\omega)
\end{equation}
with the phase shift $\pi/2$.  For linearly polarized light, the perpendicular dipole (\ref{dEM_app}) does not contribute to the ac Stark energy shift, because the time-averaged product $\langle\bm d(t)\cdot\bm E(t)\rangle_t$ vanishes, but it does for circularly polarized light, where the electric field has two perpendicular components out of phase.

For two-dimensional (2D) twisted graphene multilayers, the vectors $\bm M$ and $\bm{\mathcal M} = \mathcal M \hat{\bm z}$ are both parallel to $z$ axis in Fig.~\ref{fig:setup}.  Let us consider normal incidence of circularly polarized light 
\begin{align}  \label{E+-}
  \bm E(\omega)=(1,\pm i,0) \, E_\pm(\omega), 
\end{align}
where $E_\pm$ is the amplitude of the electric field rotating with the frequency $\omega$ either clockwise or counterclockwise in the $(x,y)$ plane.  Then the helicity vector (\ref{helicity}) also points along $z$ axis
\begin{align}
  \bm h(\omega)=\pm 2 E_\pm^2(\omega) \hat{\bm z},
\end{align}
and Eq.~(\ref{energy_shift}) becomes 
\begin{align} \label{Ua2D}
  U_a = \mp \frac12 {\rm Re}[{\mathcal M}(\omega)] \, E_\pm^2(\omega).
\end{align}
Below we describe an experimental protocol for optical control of the sign of orbital magnetization taking advantage of the helicity-dependent energy shift (\ref{Ua2D}).

%%%%%%%%%%%%%%%%%%%%%%%%%%%%%%%%%%%%%%%%%%%%%%%%%%%%%%%%%%%%%%%%%%%%%%%%%%%%%
\section{Experimental protocol}
\label{sec:experiment}
%%%%%%%%%%%%%%%%%%%%%%%%%%%%%%%%%%%%%%%%%%%%%%%%%%%%%%%%%%%%%%%%%%%%%%%%%%%%%

The Chern insulator state develops in the second-order phase transition at $T_c \approx 7.5$~K \cite{Polshyn2020}, where the system spontaneously breaks time-reversal symmetry.  Either ${\mathcal M}$ or $M$ or the Hall resistance $R_H$ can be taken as the order parameter, having positive or negative sign.  The experimentally measured sign of $R_H$ can represent topological memory.  

In the presence of circularly polarized light, the antisymmetric Stark energy shift $U_a$ breaks symmetry between the two, otherwise equienergetic, states of the system characterized by opposite signs of the orbital magnetization $\bm M$.  Thus, in thermodynamic equilibrium, the system would choose the predetermined state of lower energy $U_a$, specified by the helicity of incident light in Eq.~(\ref{Ua2D}), as opposed to choosing one of the two states randomly in the absence of circular light.  The sign of the measured Hall resistance $R_H$ indicates which state has been selected.

Conceptually, there are two possible implementations of this idea.  One option is switching at low temperature, while the sample is maintained at $T<T_c$.  In order to trigger a switch of magnetization, the energy $U_a$ in Eq.~(\ref{Ua2D}) has to be greater than the energy barrier separating the two degenerate macroscopic states with $\pm\bm M$.  This would require a rather high laser intensity.  Some numerical estimates are done in Sec.~\ref{sec:graphene}.  High laser intensity may inadvertently heat the sample (locally at the laser spot) above $T_c$ and cause other experimental complications for the low-temperature setup.  Thus, while in principle feasible, the low-temperature scenario may be difficult to implement.

Another, probably easier, experimental protocol intentionally starts at $T>T_c$, where $R_H=0$.  Then the sample is gradually cooled down below $T_c$ while being continuously illuminated by circularly polarized light.  Laser power may produce some heating, which has to be balanced by cryostat cooling.  When the sample reaches $T_c$, it undergoes a transition into the Chern insulator phase with a particular sign of $\bm M$.  Since the Landau energy at the second-order phase transition is quadratic in ${\mathcal M}$, whereas $U_a$ is linear in ${\mathcal M}$, even a weak laser intensity, represented by $E_\pm^2$ in Eq.~(\ref{Ua2D}), is sufficient to control the sign of symmetry breaking.  After the sample is cooled well below $T_c$, laser power can be turned off, but the system will preserve the sign of magnetization imposed at transition, which is manifested by the sign of the quantum Hall resistance $R_H$ at low temperature.

Effectively, the circularly polarized light acts as a ``training field'' that forces the system to choose a particular sign during symmetry breaking.  It is similar to a small training magnetic field applied to control the sign of the spontaneously developing polar Kerr effect at the superconducting transition in $\rm Sr_2RuO_4$, which breaks time-reversal symmetry \cite{Xia2006}.  In the Chern insulator, the read-out of the sign of $R_H$ can be also carried out optically by measuring the polar Kerr effect at $T<T_c$.  Yu \textit{et al.}\ \cite{Yu2021} proposed to use circularly polarized light, instead of a training magnetic field, to control the chirality of superconducting pairing, which is qualitatively similar to our proposal.

%%%%%%%%%%%%%%%%%%%%%%%%%%%%%%%%%%%%%%%%%%%%%%%%%%%%%%%%%%%%%%%%%%%%%%%%%%%%%
\begin{figure} 
\includegraphics[width=0.7\linewidth]{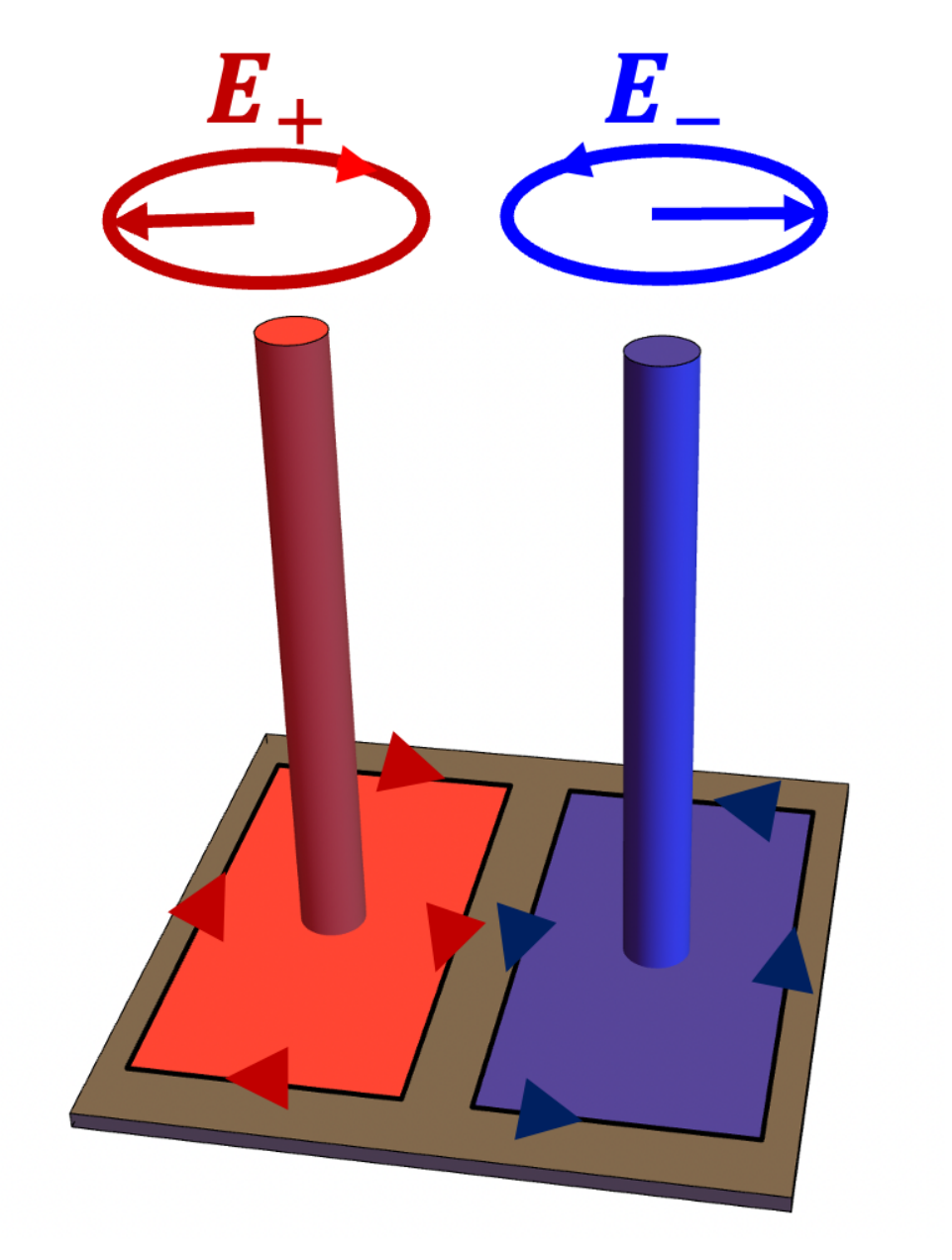}
\caption{Two laser beams of opposite circular polarization nucleate domains of opposite magnetization in the sample. A topologically-protected chiral edge channel is formed at the junction of the two domains.}
\label{fig:chiral_channel}
\end{figure}
%%%%%%%%%%%%%%%%%%%%%%%%%%%%%%%%%%%%%%%%%%%%%%%%%%%%%%%%%%%%%%%%%%%%%%%%%%%%

Moreover, by focusing two laser beams of opposite helicities on different spots in the sample, it would be possible to nucleate domains with opposite signs of orbital magnetization, as shown in Fig.~\ref{fig:chiral_channel}.  The domains would grow as temperature is lowered and produce a domain wall between them.  Such a domain wall carries topologically protected chiral gapless edge states, where electrons move in one direction only.  Since backscattering is forbidden, these metallic edge states can carry electric current without dissipation.  There is great interest in using such states, e.g.,\ as interconnects between electronic circuits to reduce dissipated power.  So far, artificially created chiral domain walls have been experimentally demonstrated in the magnetic topological insulator $\rm(Bi,Sb)_2Te_3$ doped with Cr.  The challenge of creating an inhomogeneous magnetic configuration on small scale was addressed by using the Meissner effect in a superconductor in Ref.~\cite{Rosen2017}, by scanning the sample with a magnetic tip in Ref.~\cite{Yasuda2017}, or by locally heating the sample with a laser to facilitate a magnetization flip in a magnetic field within an optically defined domain \cite{Yeats2017}.  In all of these cases, the sign of magnetization was ultimately controlled by a magnetic field acting on the Cr magnetic atoms.  In contrast, we propose an all-optical method for creating chiral domain walls in graphene multiplayers without using magnetic fields, magnetic atoms, and electric currents.

%%%%%%%%%%%%%%%%%%%%%%%%%%%%%%%%%%%%%%%%%%%%%%
\section{The ac Stark effect for a localized system with a discrete energy spectrum} 
\label{sec:discrete}
%%%%%%%%%%%%%%%%%%%%%%%%%%%%%%%%%%%%%%%%%%%%%%

For a warmup, in this section we calculate the dynamical polarizability tensor and the ac Stark energy shift for a localized system with a discrete energy spectrum, such as an atom (in contrast to a solid considered in Sec.~\ref{sec:Chern}).  We start with a single electron and then take a sum over all electrons in the system.  We set $\hbar=1$ to simplify algebra and restore it where relevant.

Substituting Eq.~(\ref{E(t)_app}) into Eq.~(\ref{E.r_app}), we get
\begin{align}
  & H = H_0 - V \,e^{-i\omega t+t\delta} -  V^\dagger \,e^{i\omega t+t\delta}, 
  \label{HV_app} \\
  & V = \frac{e r^\alpha E_\alpha(\omega)}2,\quad V^\dagger 
  = \frac{e r^\beta E_\beta^\dag(\omega)}2.
  \label{V_app}
\end{align} 
The small parameter $\delta>0$ simulates adiabatic onset of the perturbation.  The electric field in Eq.~(\ref{V_app}) can be treated as either classical or quantum field, leading to the same result in the end \cite{Delone1999,haas2006}.  In the former case, $\bm E(\omega)$ is a complex vector, so $\bm E^\dag(\omega)=\bm E^\ast(\omega)$, and the time-periodic Hamiltonian (\ref{HV_app}) implies Floquet description. In the latter case, $E_\alpha(\omega)$ and $E_\beta^\dag(\omega)$ in Eq.~(\ref{V_app}) represent the operators for stimulated annihilation and creation of a photon with the frequency $\omega$ in the presence of a laser beam.  The energy of photons should be included in $H_0$ in this case.

We seek the wavefunction as an expansion
\begin{align} \label{psi(t)-discrete}
\psi(t) = \sum_m e^{-i\varepsilon_m t} c_m(t) \, \psi^{(0)}_m
\end{align} 
in the eigenstates of the unperturbed Hamiltonian 
\begin{align}
H_0 \psi_m^{(0)} = \varepsilon_m \psi_m^{(0)}.  
\end{align} 
In this representation, the time-dependent Schr\"odinger equation 
\begin{align} \label{time-dependent}
i\dot\psi = H\psi
\end{align} 
has the form
\begin{align}  \label{schrod_c}
  -i \dot c_m = \sum_{m'\neq m} 
  \left[V_{mm'}e^{-i\omega t} + V^\dag_{mm'}e^{i\omega t}\right]
  e^{i(\varepsilon_{mm'}-i\delta)t} c_{m'},
\end{align}
where $\varepsilon_{mn}=\varepsilon_m-\varepsilon_n$, and  
\begin{equation}  \label{Vnm_app}
  V_{mn} 
  = \frac{eE_\alpha r^\alpha_{mn}}{2}, \quad
  V^\dag_{mn} = \frac{eE_\alpha^* r^\alpha_{mn}}{2}, \quad 
  r^\alpha_{mn} = \langle m|r^\alpha|n\rangle
\end{equation}
are the matrix elements of the perturbation.  The diagonal matrix elements $\bm r_{nn}=0$ vanish for the states with well-defined parity.  For a system with broken time-reversal symmetry, where $\bm r_{nm}$ is complex, interacting with circularly polarized light, where $\bm E(\omega)$ is complex, the matrix elements $|V_{mn}|\neq|V_{mn}^\dag|$ involving absorption and emission of a photon are generally not equal, which will be important below.

Suppose the wavefunction is initiated in the state $n$ of Hamiltonian $H_0$, i.e.,\ $c_m(t)\mid_{t=-\infty} = \delta_{mn}$. Then we evaluate the coefficients $c_m(t)$ using Eq.~(\ref{schrod_c}) perturbatively to the first order in $V$ and $V^\dagger$
\begin{equation}  \label{c_m_app} 
\begin{aligned}  
  & c_m(t) = c_{m-}(t) + c_{m+}(t), \\
  & c_{m-}(t) = \frac{V_{mn}e^{i(\varepsilon_{mn}-\omega -i\delta)t}}
  {\varepsilon_{mn}-\omega -i\delta}, \\
  & c_{m+}(t) = \frac{V_{mn}^\dag e^{i(\varepsilon_{mn}+\omega -i\delta)t}}
  {\varepsilon_{mn}+\omega -i\delta}. 
\end{aligned}
\end{equation}
The terms $c_{m-}$ and $c_{m+}$ represent virtual transitions with absorption or emission of a photon, respectively.  Using Eq.\ (\ref{c_m_app}), we evaluate the expectation value of the electron dipole operator (\ref{d(t)_app}) with respect to $\psi(t)$ and extract the dynamical polarizability tensor $\chi^{\alpha\beta}(\omega)$ from Eq.~$(\ref{d(omega)_app})$.  
For the initial state $n$, the corresponding contribution is
\begin{equation}  \label{chi0_app}
  \chi_n^{\alpha\beta}(\omega) = e^2 \sum_{m\neq n} \left[
  \frac{r_{nm}^\alpha r_{mn}^\beta}
  {\varepsilon_{mn}-\omega -i\delta} 
  + \frac{(r_{nm}^\alpha r_{mn}^\beta)^*}
  {\varepsilon_{mn}+\omega +i\delta} \right].
\end{equation}
The infinitesimal $i\delta$ in the denominator of the second fraction has the opposite sign because of complex conjugation. 

Now let us calculate the energy shift $\Delta\varepsilon_n$ of the state $n$ due to the ac electric field.  It can be obtained by evaluating the change $\Delta\varepsilon_n=\Delta\varepsilon_n^{(1)}+\Delta\varepsilon_n^{(2)}$ in the time-averaged expectation values of the first and the second terms in the Hamiltonian (\ref{E.r_app}) due to the electric field.  The second term $\Delta\varepsilon_n^{(2)}$ is simply the time-averaged expectation value of the dipolar energy
\begin{equation}  \label{U2_app}
  \Delta\varepsilon_n^{(2)} = - \langle\bm d(t)\cdot\bm E(t)\rangle_t 
  = -\frac12 {\rm Re} \left[
  \chi_n^{\alpha\beta}(\omega) E_\alpha^\ast(\omega)E_\beta(\omega)
  \right].
\end{equation}
The first term $\Delta\varepsilon_n^{(1)}=\Delta\langle H_0\rangle$ is the change in the expectation value of $H_0$ due to the probabilities $|c_{m\mp}|^2$ from Eq.~(\ref{c_m_app}) for occupying the states $m$.  The corresponding energy change $\varepsilon_m-\varepsilon_n\mp\omega$ results from the increase of probabilities $|c_{m\mp}|^2$ for the photon-dressed states $m$, the compensating probability decrease $-|c_{m\mp}|^2$ for the state $n$, and virtual absorption or emission of a photon at frequency $\omega$ \cite{Delone1999,haas2006}.  Thus we obtain
\begin{align}  
  \Delta\varepsilon_n^{(1)} & = \sum_{m\neq n} \left[ (\varepsilon_{mn}-\omega) \, |c_{m-}|^2
  + (\varepsilon_{mn}+\omega) \, |c_{m+}|^2 \right] \nonumber  \\
  & = {\rm Re} \sum_{m\neq n}\left[
  \frac{|V_{mn}|^2}{\varepsilon_{mn}-\omega-i\delta}
  + \frac{|V_{mn}^\dag|^2}{\varepsilon_{mn}+\omega+i\delta}
  \right] \nonumber \\
  & = \frac14 {\rm Re}\left[
  \chi_n^{\alpha\beta}(\omega) E_\alpha^\ast(\omega)E_\beta(\omega)
  \right].  \label{U1_app}
\end{align}
Adding Eqs.~(\ref{U2_app}) and (\ref{U1_app}), we express the ac Start energy shift $\Delta\varepsilon_n$ either in terms of the polarizability tensor \cite{Delone1999,haas2006}, as in Eq.~(\ref{energy_shift_def}),
\begin{equation}  \label{DE_n}
  \Delta\varepsilon_n = -\frac14 {\rm Re} \left[
  \chi_n^{\alpha\beta}(\omega) E_\alpha^\ast(\omega)E_\beta(\omega)
  \right],
\end{equation}
or in terms of the matrix elements of the perturbation \cite{Sie2015,Sie2017}
\begin{equation} \label{UV0_app}
  \Delta\varepsilon_n = - {\rm Re} \sum_{m\neq n}\left[
  \frac{|V_{mn}|^2}{\varepsilon_{mn}-\omega-i\delta}
  + \frac{|V_{mn}^\dag|^2}{\varepsilon_{mn}+\omega+i\delta}
  \right].
\end{equation}
An alternative method for derivation of $\Delta\varepsilon_n$ is presented in Appendix~\ref{sec:Alternative}.

In the case of many electrons, we label the occupied states below the Fermi energy $\varepsilon_n < E_F$ by the index $n$ and empty states with $\varepsilon_m > E_F$ by $m$.  Then the polarizability tensor in Eq.~(\ref{chi0_app}) generalizes as%
\begin{equation}  \label{chi-nm_app}
 \chi^{\alpha\beta}(\omega) =  
 \sum_{\substack{m \in \left\{\rm emp\right\}\\ n \in \left\{\rm occ\right\}}}
 \left[ \frac{e^2 \, r_{nm}^\alpha r_{mn}^\beta}
  {\varepsilon_{mn}-\omega -i\delta} 
  + \frac{e^2 \, (r_{nm}^\alpha r_{mn}^\beta)^\ast}
  {\varepsilon_{mn}+\omega +i\delta} \right].
\end{equation}
At $T\neq0$, the difference of the Fermi occupation factors $f_n(T)-f_m(T)$ should be used in Eq.~(\ref{chi-nm_app}).  We observe that Eq.~(\ref{chi-nm_app}) satisfies the standard property $\chi^{\alpha\beta}(-\omega)=\chi^{\alpha\beta}(\omega)^\ast$ of a generalized susceptibility \cite{LandauVol5}.  The ac Stark energy shift is then given by Eq.~(\ref{energy_shift_def}) with $\chi^{\alpha\beta}(\omega)$ from Eq.~(\ref{chi-nm_app}), whereas Eq.~(\ref{UV0_app}) becomes
\begin{equation} \label{UV_app}
  U = - {\rm Re} 
  \sum_{\substack{m \in \left\{\rm emp\right\}\\ n \in \left\{\rm occ\right\}}}
  \left[ \frac{|V_{mn}|^2}{\varepsilon_{mn}-\omega-i\delta}
  + \frac{|V_{mn}^\dag|^2}{\varepsilon_{mn}+\omega+i\delta} \right].
\end{equation}
Beware that generally $|V_{mn}|^2\neq|V_{mn}^\dag|^2$.

The symmetric and antisymmetric parts of the polarizability tensor $\chi^{\alpha\beta}(\omega)$ in Eq.~(\ref{chi-nm_app}) are
\begin{align}
   & \chi^{\alpha\beta}_s(\omega) = 2e^2 
   \sum_{\substack{m \in \left\{\rm emp\right\}\\ n \in \left\{\rm occ\right\}}}
   \frac{\varepsilon_{mn}\,{\rm Re}(r^\alpha_{nm} r^\beta_{mn})}
   {\varepsilon_{nm}^2-(\hbar\omega+i\delta)^2}, \label{chi_s_app} \\
    & \chi^{\alpha\beta}_a(\omega) = 2e^2 
   \sum_{\substack{m \in \left\{\rm emp\right\}\\ n \in \left\{\rm occ\right\}}}
   \frac{i\hbar\omega \,{\rm Im}(r^\alpha_{nm} r^\beta_{mn})}
   {\varepsilon_{nm}^2-(\hbar\omega+i\delta)^2},
   \label{chi_a_app}
\end{align}
where we restored $\hbar$ by replacing $\omega\to\hbar\omega$.  The antisymmetric polarizability $\chi^{\alpha\beta}_a(\omega)$ is only permitted for systems with broken time-reversal symmetry at a nonzero $\omega$.   It can be represented via Eq.~(\ref{chi-M}) in terms of the dual vector 
\begin{equation}  \label{M_app}
  \bm{\mathcal M}(\omega) =  - e^2 
  \sum_{m \in {\rm emp}, \: n \in {\rm occ}}
  \frac{\hbar\omega \,{\rm Im}(\bm r_{nm} \times \bm r_{mn})}
  {\varepsilon_{nm}^2-(\hbar\omega+i\delta)^2}.
\end{equation}
The antisymmetric energy shift $U_a$ in Eq.~(\ref{energy_shift}) is obtained as the scalar product of the axial time-reversal-odd vector $\bm{\mathcal M}(\omega)$ from Eq.~(\ref{M_app}) and the helicity $\bm h(\omega)$ of circularly polarized light from Eq.~(\ref{helicity}).

In the high-frequency limit, the leading terms in Eq.~(\ref{M_app}) are
\begin{align}  
  \bm{\mathcal M}(\omega) \approx {} & \frac{e^2}{\hbar\omega} 
  \sum_{m \in {\rm emp}, \: n \in {\rm occ}}
  {\rm Im}(\bm r_{nm} \times \bm r_{mn})
  \label{M-omega>>1} \\
  & + \frac{e^2}{(\hbar\omega)^3} 
  \sum_{m \in {\rm emp}, \: n \in {\rm occ}}
  \varepsilon_{nm}^2\,{\rm Im}(\bm r_{nm} \times \bm r_{mn}).
  \notag
\end{align}
The coefficient in the first term vanishes, because it can be written as $\langle n|\bm r \times \bm r|n\rangle=0$ due to the completeness relation.  In the second term, we use the operator relation
\begin{align} \label{v-r}
  \hbar\hat{\bm v} = i[H,\bm r]
\end{align}
to connect the matrix elements of velocity and position operators
\begin{align} \label{v-r-mn}
  \hbar v_{nm}^\alpha = i\varepsilon_{nm}\,r_{nm}^\alpha.
\end{align}
Substituting Eq.~(\ref{v-r-mn}) into Eq.~(\ref{M-omega>>1}) and using the completeness relation, we find
\begin{align}  \label{M-1/omega^3}
  \bm{\mathcal M}(\omega) \approx \frac{e^2}{\hbar\omega^3} \, 
  {\rm Im}\sum_{n \in {\rm occ}} \langle n|\hat{\bm v} \times \hat{\bm v}|n\rangle.
\end{align}
Suppose an atom is subject to an external magnetic field $\bm B=\bm\nabla\times\bm A$, where $\bm A$ is the vector potential.  Then the velocity operator is
\begin{align} \label{v-A}
  \hat{\bm v} = \frac{\hat{\bm p} - e\bm A}{m_e} ,
\end{align}
where $\hat{\bm p}$ is the electron momentum operator, and $m_e$ is the electron mass.  Substituting Eq.~(\ref{v-A}) into Eq.~(\ref{M-1/omega^3}), we find
\begin{align}  \label{M-B}
  \bm{\mathcal M}(\omega) \approx \frac{e^3}{m_e^2 \omega^3} \, N \bm B,
\end{align}
where $N$ is the number of electrons in the atom.  Substituting Eq.~(\ref{M-B}) into Eq.~(\ref{energy_shift}), we obtain the energy shift of the atom
\begin{align}  \label{U-B}
  U_a \approx - \frac{e^3}{4 m_e^2 \omega^3} \, N \bm B\cdot\bm h(\omega)
  = -\bm M \cdot \bm B,
\end{align}
where
\begin{align}  \label{M-h}
  \bm M \approx \frac{e^3}{4 m_e^2 \omega^3} \, N \bm h(\omega)
\end{align}
is the static magnetic moment induced in the atom by circularly polarized light with helicity $\bm h(\omega)$.  Equation (\ref{M-h}) gives the high-frequency limit of the inverse Faraday effect \cite{vanderZiel1965,Pershan1966}.  In this limit, the response of electrons in an atom is similar to that of free electrons  \cite{Pitaevskii1961}.

%%%%%%%%%%%%%%%%%%%%%%%%%%%%%%%%%%%%%%%%%%%%%%
\section{The ac Stark effect for a Chern insulator}
\label{sec:Chern}
%%%%%%%%%%%%%%%%%%%%%%%%%%%%%%%%%%%%%%%%%%%%%%

Now let us generalize the results of Sec.\ \ref{sec:discrete} for electrons in a crystal.  The electronic Bloch states are labeled by the discrete band index $n$ and continuous quasimomentum $\bm k$ (simply called momentum in the rest of the paper for shortness).  The wavefunctions for the energies $\varepsilon_n(\bm k)$ have the Bloch form $\psi_{n,\bm k} = e^{i\bm k\cdot\bm r} u_{n,\bm k}(\bm r)$, where $u_{n,\bm k}(\bm r)$ is periodic in $\bm r$.  We do not write the spin index explicitly, so summation over the two spin orientations should be added to the final results.  

The dipolar coupling to the electric field in Eq.~(\ref{E.r_app}) breaks translational symmetry of crystal, so the momentum $\bm k$ is generally not conserved.  Thus, in contrast to Eq.~(\ref{psi(t)-discrete}), the following expansion of the wavefunction is more practical \cite{Bauer2020}
\begin{align} \label{psi-k(t)}
  \psi(t) = e^{i\bm r\cdot\tilde{\bm k}(t)} \sum_m c_{m,\bm k}(t) \, 
  e^{-i\int\limits^t dt' \varepsilon_m[\tilde{\bm k}(t')]} \,
  u_{m,\tilde{\bm k}(t)}(\bm r).
\end{align} 
Here $\tilde{\bm k}(t)$ is the time-dependent momentum of the electron, which, by definition, satisfies Newton's equation of motion
\begin{align} \label{Newton}
  \frac{d\tilde{\bm k}(t)}{dt}=e\bm E(t),
\end{align} 
whereas $\bm k=\langle\tilde{\bm k}(t)\rangle_t$ is its time-averaged value, noting that $\langle\bm E(t)\rangle_t=0$ in Eq.~(\ref{E(t)_app}).  Substituting Eq.~(\ref{psi-k(t)}) into the time-dependent Schr\"odinger equation  Eq.~(\ref{time-dependent}), we obtain an analog of Eq.~(\ref{schrod_c})
\begin{align} \label{c(t)-k(t)}
  -i\dot c_{m,\bm k} = i\sum_{m'} eE_\alpha 
  \langle u_{m,\tilde{\bm k}}|\partial^\alpha u_{m',\tilde{\bm k}}\rangle
  e^{i\int\limits^t dt' \varepsilon_{mm'}(\tilde{\bm k})} c_{m',\bm k}.
\end{align} 
Here $\partial^\alpha = \partial/\partial k_\alpha$ denotes the momentum derivative, and we omitted the time argument of $\tilde{\bm k}(t)$ to shorten the notation.  Equation (\ref{c(t)-k(t)}) reflects the notion that matrix elements of the coordinate operator $\bm r$ in a crystal are expressed in terms of the Berry connections \cite{BlountBook1962,LandauBook}
\begin{align} \label{connections}
  r_{nm}^\alpha(\bm k) = i\langle u_{n,\bm k}|\partial^\alpha u_{m,\bm k}\rangle
  = -i\langle\partial^\alpha  u_{n,\bm k}|u_{m,\bm k}\rangle
  = r_{mn}^{\alpha *}(\bm k).
\end{align} 

The term with $m'=m$ in Eq.~(\ref{c(t)-k(t)}) produces the Berry phase factor
\begin{align} \label{c-Berry}
  c_{m,\bm k}(t) = \tilde c_{m,\bm k}(t) \, e^{i\phi_{m,\bm k}(t)},
\end{align}
where the Berry phase $\phi_{m,\bm k}(t)$ is defined by the equation
\begin{align} \label{Berry-phase}
  \frac{d\phi_{m,\bm k}(t)}{dt}= 
  i \langle u_{m,\tilde{\bm k}}|\partial^\alpha u_{m,\tilde{\bm k}}\rangle 
  \frac{d\tilde k_\alpha}{dt}.
\end{align}
When $\tilde{\bm k}(t)$ moves along a small closed loop around its mean value $\bm k$, the accumulated Berry phase $\Delta\phi_{m,\bm k}$ is expressed as an integral over the enclosed area in the momentum space
\begin{align} \label{D-phi}
  \Delta\phi_{m,\bm k} = \epsilon^{\alpha\beta\gamma} 
  \int dk_\alpha \, dk_\beta \, \Omega_{m,\gamma}(\bm k)
\end{align}
of the Berry curvature vector
\begin{equation}  \label{Berry-curvature}
  \Omega_{m,\gamma}(\bm k) = i\epsilon_{\alpha\beta\gamma}\, 
  \langle \partial^\alpha u_{m,\bm k}|\partial^\beta u_{m,\bm k} \rangle.
\end{equation}

In 2D case, the Berry curvature vector $\bm\Omega_n(\bm k)=\hat{\bm z}\,\Omega_n(\bm k)$ points along $z$ axis.  From the equation of motion (\ref{Newton}) with the rotating electric field (\ref{E+-}), we find that the momentum $\tilde{\bm k}(t)$ moves clockwise or counterclockwise on a circle of radius $eE_\pm/\omega$.  The circle encloses the area $\pi(eE_\pm/\omega)^2$ in the momentum space.  Thus, by Eq.~(\ref{D-phi}), the Berry phase increases by $\Delta\phi_{n,\bm k}=\pm\Omega_{n}(\bm k)\,\pi(eE_\pm/\omega)^2$ per cycle.  The time rate of the Berry phase increase is $\Delta\phi_{n,\bm k}\omega/2\pi$, thus, on average, the Berry phase increases in time as
\begin{align} \label{Berry(t)}
  \phi_{n,\bm k}(t) = \pm t\,\Omega_n(\bm k)\,\frac{(eE_\pm)^2}{2\omega}.
\end{align}
The linearly increasing in time Berry phase (\ref{Berry(t)}) results in the energy shift of the state $|n,\bm k\rangle$ in Eq.~(\ref{c-Berry})
\begin{align} \label{DE-Berry}
  \Delta\varepsilon_n^{\rm intra}(\bm k)= \pm\Omega_n(\bm k)\,\frac{(eE_\pm)^2}{2\omega}.
\end{align}
We label this term as ``intra'', because it originates from the intraband diagonal matrix element in Eq.~(\ref{connections}).  In 3D notation, Eq.~(\ref{DE-Berry}) has the form \begin{align} \label{DE-intra-3D}
  \Delta\varepsilon_n^{\rm intra}(\bm k)= 
  - \frac{e^2 \, \bm h(\omega)\cdot\bm\Omega_n(\bm k)}{4\omega}.
\end{align}
This term is only present for circularly or elliptically polarized light and absent for linear polarization, because the electron trajectory does not enclose an area in momentum space in the latter case.

Now let us substitute Eq.~(\ref{c-Berry}) into Eq.~(\ref{c(t)-k(t)}), which eliminates the term with $m'=m$:
\begin{align} \label{tilde-c(t)-k}
  -i\dot{\tilde c}_{m,\bm k} \approx i\sum_{m'\neq m} eE_\alpha 
  \langle u_{m,\bm k}|\partial^\alpha u_{m',\bm k}\rangle
  e^{i\varepsilon_{mm'}(\bm k)t} \tilde c_{m',\bm k}.
\end{align}
Here $\tilde{\bm k}(t)$ is replaced by $\bm k$, and the Berry phases $\phi_m$ and $\phi_{m'}$ are omitted in the exponential factor, because they produce corrections of a higher order in the small electric field $E$.  Equation (\ref{tilde-c(t)-k}) has the same structure as Eqs.~(\ref{schrod_c}) and (\ref{Vnm_app}).  Thus, using Eqs.~(\ref{chi_a_app}), (\ref{chi-M}), and (\ref{energy_shift}), we obtain the antisymmetric interband contribution to the energy shift of the state $|n,\bm k\rangle$
\begin{align} \label{DE-inter-3D}
  \Delta\varepsilon_{n,a}^{\rm inter}(\bm k)= 
  - \frac{e^2}{4} \, \bm h(\omega)\cdot 
  {\sum_{m\neq n} \rm Re} \frac{\bm\Xi_{nm}(\bm k)\,\omega}
  {\varepsilon_{nm}^2(\bm k)-(\omega+i\delta)^2}.
\end{align}
Here $\varepsilon_{nm}(\bm k)=\varepsilon_n(\bm k)-\varepsilon_m(\bm k)$, and the dual vector $\bm\Xi_{nm}(\bm k)$ is defined in terms of the antisymmetric tensor of the interband Berry connections (\ref{connections}) with $m\neq n$
\begin{equation}  \label{Im-rr_app}
  {\rm Im }\left[r^\alpha_{nm}(\bm k)\,r^\beta_{mn}(\bm k)\right] 
  = - \frac12 \, \epsilon^{\alpha\beta\gamma} \, \Xi_{nm,\gamma}(\bm k).
\end{equation}
Combining the intra- and interband contributions (\ref{DE-intra-3D}) and (\ref{DE-inter-3D}), we get the total antisymmetric energy shift of the state $|n,\bm k\rangle$
\begin{align}  \label{DE-total-3D}
  \Delta\varepsilon_{n,a}(\bm k) =  -\frac14 \, \bm h(\omega) 
  \cdot {\rm Re}\,\bm{\mathcal M}_{n,\bm k}(\omega)
\end{align}
with
\begin{align}  \label{M_n}
  \bm{\mathcal M}_{n,\bm k}(\omega) = e^2 \left[ \sum_{m\neq n} 
  \frac{\bm\Xi_{nm}(\bm k)\,\hbar\omega}{\varepsilon_{nm}^2(\bm k)-(\hbar\omega+i\delta)^2} 
  + \frac{\bm\Omega_n(\bm k)}{\hbar\omega} \right],
\end{align}
where we restored $\hbar$ by replacing $\omega\to\hbar\omega$.

At low frequency, $\bm{\mathcal M}_{n,\bm k}(\omega) \propto \bm\Omega_n(\bm k)/\hbar\omega$ is dominated by the second term in Eq.~(\ref{M_n}) involving the Berry curvature of the state $|n,\bm k\rangle$.  Using this limit in Eq.~(\ref{dEM_app}) and transforming to the time domain, we obtain the time derivative of the dipole moment
\begin{equation}  \label{v-d-nk}
  \dot{\bm d}_{n,\bm k}(t) = - \frac{e^2}{\hbar} 
  \bm E(t)\times\bm\Omega_n(\bm k) = e \bm v_{n,\bm k}(t),
\end{equation} 
which represents the anomalous Hall velocity $\bm v_{n,\bm k}$ carried by the state $|n,\bm k\rangle$.  Equation (\ref{v-d-nk}) implies that, for a nonzero Berry curvature, a time-independent electric field $\bm E$ induces a steady transverse velocity $\bm v$, so the dipole moment $\bm d\propto t$ increases linearly in time.  This is an important difference between electric polarizations of a solid and a localized system studied in Sec.~\ref{sec:discrete}, where the induced dipole moment is always finite.

Equation (\ref{M_n}) can be rewritten as
\begin{align}  \label{M_n-Xi}
  \bm{\mathcal M}_{n,\bm k}(\omega) = \frac{e^2}{\hbar\omega} \sum_{m\neq n} 
  \frac{\varepsilon_{nm}^2(\bm k)\,\bm\Xi_{nm}(\bm k)}
  {\varepsilon_{nm}^2(\bm k)-(\hbar\omega+i\delta)^2}
\end{align}
using the sum rule
\begin{align}  \label{Xi-Omega-sum}
  \bm\Omega_n(\bm k) & = \sum_{m\neq n}\bm\Xi_{nm}(\bm k) \\
  & = i\epsilon_{\alpha\beta\gamma}\, 
  \sum_{m\neq n} \langle\partial^\alpha u_{n,\bm k}|u_{m,\bm k}\rangle
  \langle u_{m,\bm k}|\partial^\beta u_{n,\bm k} \rangle .
  \nonumber
\end{align}
Similarly to Eq.~(\ref{M-1/omega^3}), it follows from Eq.~(\ref{M_n-Xi}) that $\bm{\mathcal M}_n(\omega) \propto 1/\omega^3$ at high frequency, so the transverse current is strongly suppressed because of cancellation between intra- and interband terms in Eq.~(\ref{M_n}).

To generalize Eq.~(\ref{M_n-Xi}) for many electrons in 2D, we introduce a sum over the occupied and empty bands and include the integral $\sum_{\bm k} \to \mathcal A \int d^2k/(2\pi)^2$, where $\mathcal A$ is the area of the sample.  All dual vectors $\bm{\mathcal M}_n(\omega)$, $\bm\Xi_{nm}(\bm k)$, and $\bm\Omega_n(\bm k)$ are perpendicular to the 2D plane and parallel to $z$ axis.  Thus, Eq.~(\ref{M_n-Xi}) becomes
\begin{align}  \label{M-2D}
 \frac{{\mathcal M}_b(\omega)}{\mathcal A} = \int\frac{d^2k}{(2\pi)^2}
 \sum_{\substack{m \in \left\{\rm emp\right\}\\ n \in \left\{\rm occ\right\}}}
 \frac{e^2}{\hbar\omega} \frac{\varepsilon_{nm}^2(\bm k)\,\Xi_{nm}(\bm k)}
  {\varepsilon_{nm}^2(\bm k)-(\hbar\omega+i\delta)^2}.
\end{align}
Here the subscript $b$ indicates that ${\mathcal M}_b(\omega)$ represents a contribution from the bulk electronic states, as opposed to the edge states discussed in Sec.~\ref{sec:edge}.

In the limit of low frequency $\omega$, Eq.~(\ref{M-2D}) gives
\begin{align}  \label{M-low-omega}
  \frac{\mathcal M_b(\omega)}{\mathcal A} 
  \approx \frac{e^2}{2\pi}\,\frac{C}{\hbar\omega},
\end{align}
where
\begin{align} \label{Chern_app}
  C = \sum_{n \in {\rm occ}} \int\frac{d^2k}{2\pi} \, \Omega_n(\bm k)
\end{align}
is the integer topological Chern number.  Substituting Eq.~(\ref{M-low-omega}) into Eq.~(\ref{dEM_app}) and transforming to the time domain, we recover the integer quantum Hall effect for a Chern insulator
\begin{equation}  \label{QHE_b}
  \bm j_b = - C\,\frac{e^2}{h} \left[\bm E\times\hat{\bm z}\right],
\end{equation} 
for is the 2D bulk current density
\begin{align} \label{j_b-d}
  \bm j_b = \frac{\dot{\bm d}_b}{\mathcal A}
  = \frac{e\bm v_b}{\mathcal A}.
\end{align}
In the opposite limit of high frequency, ${\mathcal M}_b \propto 1/\omega^3$, so the Hall current is suppressed as $j_b\propto1/\omega^2$.

Further simplification can be achieved for a two-band insulator consisting of only the occupied valence and empty conduction bands, labeled as $v$ and $c$.  In this case, the sum in Eq.~(\ref{Xi-Omega-sum}) contains only one term, so 
\begin{equation}  \label{Omega=Xi}
  \Omega_v(\bm k) = \Xi_{vc}(\bm k).
\end{equation} 
Then Eq.~(\ref{M-2D}) becomes
\begin{equation}  \label{M_Berry_app}
   \frac{{\mathcal M}_b(\omega)}{\mathcal A} = \frac{e^2}{\hbar\omega} 
   \int\frac{d^2k}{(2\pi)^2}
   \frac{\varepsilon_{cv}^2(\bm k) \, \Omega_v(\bm k)}
   {\varepsilon_{cv}^2(\bm k)-(\hbar\omega+i\delta)^2}.
\end{equation}
The polarization vector ${\mathcal M}_b(\omega)$ is expressed in Eq.~(\ref{M_Berry_app}) for a two-band model as an integral involving the Berry curvature.  The static orbital magnetization $M$ of a Chern insulator is also expressed as a different integral involving the Berry curvature \cite{Niu2005,Thonhauser2005}.  Thus, $M$ and ${\mathcal M}_b(\omega)$ are not equivalent, but their signs are both determined by the sign of the Berry curvature $\Omega_v(\bm k)$.  An application of Eq.~(\ref{M_Berry_app}) to the Haldane model on the honeycomb lattice is illustrated in Appendix \ref{appendix:Haldane}.

Substituting Eq.~(\ref{M_Berry_app}) into Eq.~(\ref{energy_shift}), we finally obtain the antisymmetric ac Stark energy shift due to the bulk states for a 2D two-band Chern insulator
\begin{align}
    \frac{U_a^{(b)}}{\mathcal A} = & - \frac{e^2}{4\hbar\omega}
    \,{\rm Im}[\bm E^\ast(\omega)\times \bm E(\omega)]_z 
    \label{stark_two_bands}\\
    &  \times {\rm Re} \int\frac{d^2k}{(2\pi)^2}
    \frac{\varepsilon_{cv}^2(\bm k) \, \Omega_v(\bm k)}
    {\varepsilon_{vc}^2(\bm k)-(\hbar\omega+i\delta)^2}. \nonumber
\end{align}
While the quantum Hall effect for a Chern insulator was much discussed in the literature, the Stark energy shift (\ref{stark_two_bands}) in a circularly polarized electric field was not discussed, to the best of our knowledge.

%%%%%%%%%%%%%%%%%%%%%%%%%%%%%%%%%%%%%%%%%%%%%%
\section{Relation with the ac Hall effect}
\label{sec:acHall}
%%%%%%%%%%%%%%%%%%%%%%%%%%%%%%%%%%%%%%%%%%%%%%

It is well known that the electric field (\ref{E(t)_app}) can be introduced into the Hamiltonian of the system using different gauges.  In Eq.~(\ref{E.r_app}), we use the gauge where the scalar potential is $A_0(t)=-\bm r\cdot\bm E(t)$, whereas the vector potential is zero $\bm A=0$.  Alternatively, we can use the gauge where $A_0=0$ and
\begin{equation}  \label{A(t)}
  \bm A(t) = \frac12 \left[\bm A(\omega)\, e^{-i\omega t} 
  + \bm A(-\omega)\, e^{i\omega t}\right],
\end{equation}
so that 
\begin{align}  \label{E=-dA/dt}
  \bm E(t)=-\partial_t\bm A(t), \qquad \bm E(\omega)=i\omega\bm A(\omega).
\end{align}
The calculated ac Stark energy shift should be the same in either gauge, as it was shown explicitly for the symmetric part $U_s$ in Ref.~\cite{Kobe1983}.  In this Section, we rederive the antisymmetric part $U_a$ using the $\bm A$ gauge for a 2D Chern insulator and establish connection with the ac Hall effect.

In the presence of the vector potential $\bm A$, the Hamiltonian of the system 
\begin{align}  \label{H(k-A)}
  H(t) = H_0[\bm k-e\bm A(t)]
\end{align}
is obtained by the substitution $\bm k \to \tilde{\bm k}(t)$, where
\begin{align}  \label{k-A}
  \tilde{\bm k}(t) = \bm k-e\bm A(t)
\end{align}
is the time-dependent momentum already introduced in Sec.~\ref{sec:Chern} and satisfying Eq.~(\ref{Newton}), given Eq.~(\ref{E=-dA/dt}).  One approach is to 
represent the wavefunction $\psi(t)$ as a superposition (\ref{psi-k(t)}) of the instantaneous eigenstates of the Hamiltonian $H(t)$ \cite{Bauer2020}, thus reproducing the results of Sec.~\ref{sec:Chern}.  In particular, the intraband contribution (\ref{DE-intra-3D}) to the energy shift $\Delta\varepsilon_n^{\rm intra}(\bm k)$ can be obtained as the time-averaged value of the diagonal coefficient with $m'=m$ in Eq.~(\ref{c(t)-k(t)})
\begin{align}  
  & e\langle E_\alpha(t) r_{nn}^\alpha[\bm k-e\bm A(t)] \rangle_t 
  \approx -e^2 \partial^\beta r_{nn}^\alpha(\bm k) 
  \langle E_\alpha(t)A_\beta(t)\rangle_t \notag \\
  & = -\frac{e^2}{2\omega}\,
  {\rm Im}\left[E_\alpha^\ast(\omega)E_\beta(\omega)\right]
  \langle \partial^\beta u_{n,\bm k}|\partial^\alpha u_{n,\bm k} \rangle \notag \\
  & = - \frac{e^2 \, \bm h(\omega)\cdot\bm\Omega_n(\bm k)}{4\omega} 
  = \Delta\varepsilon_n^{\rm intra}(\bm k). 
  \label{m'=m}
\end{align}

Another approach is to expand the Hamiltonian (\ref{H(k-A)}) in the powers of $\bm A(t)$.  To the first order,
\begin{align}  \label{H-A}
  H(t) \approx H_0 - e \bm A(t)\cdot \hat{\bm v},
\end{align}
where 
\begin{align}  \label{v=dH/dk}
  \hat{\bm v}(\bm k) = \frac{\partial H_0}{\partial\bm k} 
\end{align}
is the operator of total velocity, which is the sum of velocity operators for all electrons in the system.  We use the hat to distinguish the operator $\hat{\bm v}$ from its expectation value $\bm v(t)$
\begin{equation}  \label{<v>}
  \langle\hat{\bm v}\rangle = \bm v(t) 
  = \frac12 \left[\bm v(\omega)\, e^{-i\omega t} 
  + \bm v(-\omega)\, e^{i\omega t}\right],
\end{equation}
which develops in response to the vector potential (\ref{A(t)}).  The second-order, diamagnetic term $(e^2/2)\,A_\alpha A_\beta \, \partial^\alpha \partial^\beta H_0$ in the expansion (\ref{H-A}) is symmetric and, thus, not relevant for the antisymmetric response, so it is omitted.   However, this term is important for ensuring gauge invariance of the symmetric response \cite{Kobe1983}.

Using a linear response theory similar to Sec.~\ref{sec:Chern} for the Hamiltonian (\ref{H-A}) and focusing on its antisymmetric part (\ref{chi_a_app}), we find the current density $\bm j_b(\omega)$ carried by the bulk states (labeled by the subscript $b$)
\begin{equation}  \label{j_b-alpha}
   j_b^\alpha(\omega) = \frac{e v_b^\alpha(\omega)}{\mathcal A}
   = \sigma_{H,b}^{\alpha\beta}(\omega) \, E_\beta(\omega),
\end{equation}
where the bulk ac Hall conductivity tensor is
\begin{align}  \label{sigma_H-v}
  & \sigma_{H,b}^{\alpha\beta} (\omega) = 2e^2 
  \sum_{\substack{m \in \left\{\rm emp\right\}\\ n \in \left\{\rm occ\right\}}}
  \int\frac{d^2k}{(2\pi)^2}
  \frac{\hbar \, {\rm Im }\left[v^\alpha_{nm}(\bm k)v^\beta_{mn}(\bm k)\right]}
  {\varepsilon_{nm}^2(\bm k)-(\hbar\omega+i\delta)^2} \\
  & = \frac{2e^2}{\hbar} 
  \sum_{\substack{m \in \left\{\rm emp\right\}\\ n \in \left\{\rm occ\right\}}}
  \int\frac{d^2k}{(2\pi)^2}
  \frac{\varepsilon_{nm}^2(\bm k) \, {\rm Im }\left[r^\alpha_{nm}(\bm k)r^\beta_{mn}(\bm k)\right]}
  {\varepsilon_{nm}^2(\bm k)-(\hbar\omega+i\delta)^2}.
  \label{sigma_H-r}
\end{align}
Here Eq.~(\ref{v-r-mn}) was used to transform Eq.~(\ref{sigma_H-v}) into Eq.~(\ref{sigma_H-r}).
Substituting Eq.~(\ref{Im-rr_app}) into Eq.~(\ref{sigma_H-r}), we express the bulk ac Hall conductivity tensor in 2D
\begin{equation}  \label{Hall-tensor}
  \sigma_{H,b}^{\alpha\beta} (\omega) = \epsilon^{\alpha\beta}\,\sigma_H^{(b)}(\omega),
\end{equation}
in terms of the bulk ac Hall conductivity
\begin{align}  \label{sigma_H-Xi}
  \sigma_H^{(b)}(\omega) = - \frac{e^2}{\hbar} 
  \sum_{\substack{m \in \left\{\rm emp\right\}\\ n \in \left\{\rm occ\right\}}}
  \int\frac{d^2k}{(2\pi)^2}
  \frac{\varepsilon_{nm}^2(\bm k) \, \Xi_{nm}(\bm k)}
  {\varepsilon_{nm}^2(\bm k)-(\hbar\omega+i\delta)^2}.
\end{align}
In the dc limit $\omega=0$, Eq.~(\ref{sigma_H-Xi}) gives the integer quantum Hall conductivity consistent with Eq.~(\ref{QHE_b})
\begin{equation}  \label{sigma_H-dc}
  \sigma_H^{(b)}(\omega=0) = -C \, \frac{e^2}{h},
\end{equation}
where $C$ is the integer topological Chern number from Eqs.~(\ref{Chern_app}) and (\ref{Xi-Omega-sum}).  For a two-band Chern insulator, Eq.~(\ref{sigma_H-Xi}) reduces to
\begin{equation}  \label{sigma_H-Berry}
  \sigma_H^{(b)}(\omega) = - \frac{e^2}{\hbar} \int\frac{d^2k}{(2\pi)^2}
  \frac{\varepsilon_{vc}^2(\bm k) \, \Omega_v(\bm k)}
  {\varepsilon_{vc}^2(\bm k)-(\hbar\omega+i\delta)^2},  
\end{equation}
in agreement with the earlier calculations \cite{YuguiYao2004,Yakovenko2008} of the bulk ac Hall conductivity for a two-band model.  

Now we calculate the energy shift $U_a$ due to the antisymmetric response to the perturbation $\bm A$ in the Hamiltonian (\ref{H-A}) by analogy with Eqs.~(\ref{U2_app}), (\ref{DE_n}), and (\ref{energy_shift_def})
\begin{equation}  \label{DUAv_app}
  U_a = - \frac e2 \langle\bm v(t)\cdot\bm A(t)\rangle_{t,a}.
\end{equation}
Using Eqs.~(\ref{E=-dA/dt}), (\ref{j_b-alpha}) and (\ref{Hall-tensor}), we obtain the bulk contribution to $U_a$
\begin{align}  
  \frac{U_a^{(b)}}{\mathcal A} & = \frac14 \, {\rm Re} \, 
  \frac{\sigma_{H,b}^{\alpha\beta}(\omega) E_\alpha^\ast(\omega) E_\beta(\omega)}
  {i\omega} \nonumber \\
  & =   \frac{{\rm Re} \, \sigma_H^{(b)}(\omega)}{4\omega} \, h_z(\omega).
  \label{Uab_app}
\end{align}
For linearly polarized light, the antisymmetric energy shift $U_a$ vanishes in Eq.~(\ref{DUAv_app}), because the vector potential $\bm A \| \bm E$ is parallel to $\bm E$, whereas the Hall current $\bm v \!\perp\! \bm E$ is perpendicular to $\bm E$, so $\bm v \!\perp\! \bm A$.  In contrast, for circularly polarized light, the vector potential $\bm A\!\perp\!\bm E$ is perpendicular to $\bm E$ in Eq.~(\ref{E=-dA/dt}) and thus parallel to the Hall current $\bm v \| \bm A$, which produces a nonzero term $U_a^{(b)}$ proportional the helicity of light in Eq.~(\ref{Uab_app}).

The two expressions for $U_a$, presented by Eq.~(\ref{Uab_app}) in terms of the Hall conductivity (\ref{sigma_H-Xi}) and by Eq.~(\ref{energy_shift}) in terms of the polarizability vector (\ref{M-2D}), agree with each other, given that the polarizability vector and the Hall conductivity are related as 
\begin{align}  \label{M=sigma_H/omega}
  \frac{{\mathcal M}_b(\omega)}{\mathcal A} = -\frac{\sigma_H^{(b)}(\omega)}{\omega}.  
\end{align}
The relation (\ref{M=sigma_H/omega}) is naturally expected, because the electric current density is the time derivative of the electric dipole moment in Eq.~(\ref{j_b-d}).  In the high-frequency limit, the Hall conductivity tensor from Eq.~(\ref{sigma_H-v}) decreases as $1/\omega^2$ and is proportional the expectation value of the commutator of velocity operators \cite{Shastry1993}
\begin{align}  \label{sigma_H-high-omega}
  \sigma_{H,b}^{\alpha\beta} (\omega) \approx -\frac{2e^2}{\hbar\omega^2} 
  \sum_{n \in \left\{\rm occ\right\}}
  \int\frac{d^2k}{(2\pi)^2}
  {\rm Im }\langle n|\!
  \left[\hat v^\alpha(\bm k),\hat v^\beta(\bm k)\right]\! |n\rangle
\end{align}
The corresponding polarization vector ${\mathcal M}_b(\omega)$ in Eq.~(\ref{M=sigma_H/omega}) decreases as $1/\omega^3$ in agreement with Eq.~(\ref{M-1/omega^3}).

Thus we have demonstrated that either representation of coupling to the electric field, in terms of $\bm E\cdot\bm r$ in Eq.~(\ref{E.r_app}) or $\bm A\cdot\hat{\bm v}$ in Eq.~(\ref{H-A}), produces the same result for the antisymmetric ac Stark energy shift $U_a$.  To achieve the agreement, it is crucial to take into account the intraband contribution (\ref{DE-intra-3D}) or (\ref{m'=m}) to the energy shift due to the Berry curvature.  The agreement was demonstrated earlier for the symmetric part $U_s$ in Ref.~\cite{Kobe1983} taking into account the quadratic diamagnetic term in the expansion (\ref{H-A}).

%%%%%%%%%%%%%%%%%%%%%%%%%%%%%%%%%%%%%%%%%%%%%%
\section{Contribution from the edge states}
\label{sec:edge}
%%%%%%%%%%%%%%%%%%%%%%%%%%%%%%%%%%%%%%%%%%%%%%

%%%%%%%%%%%%%%%%%%%%%%%%%%%%%%%%%%%%%%%%%%%%%%%%%%%%%%%%%%%%%%%%%%%%%%%%%%%%%
\begin{figure}
	\includegraphics[width=\linewidth]{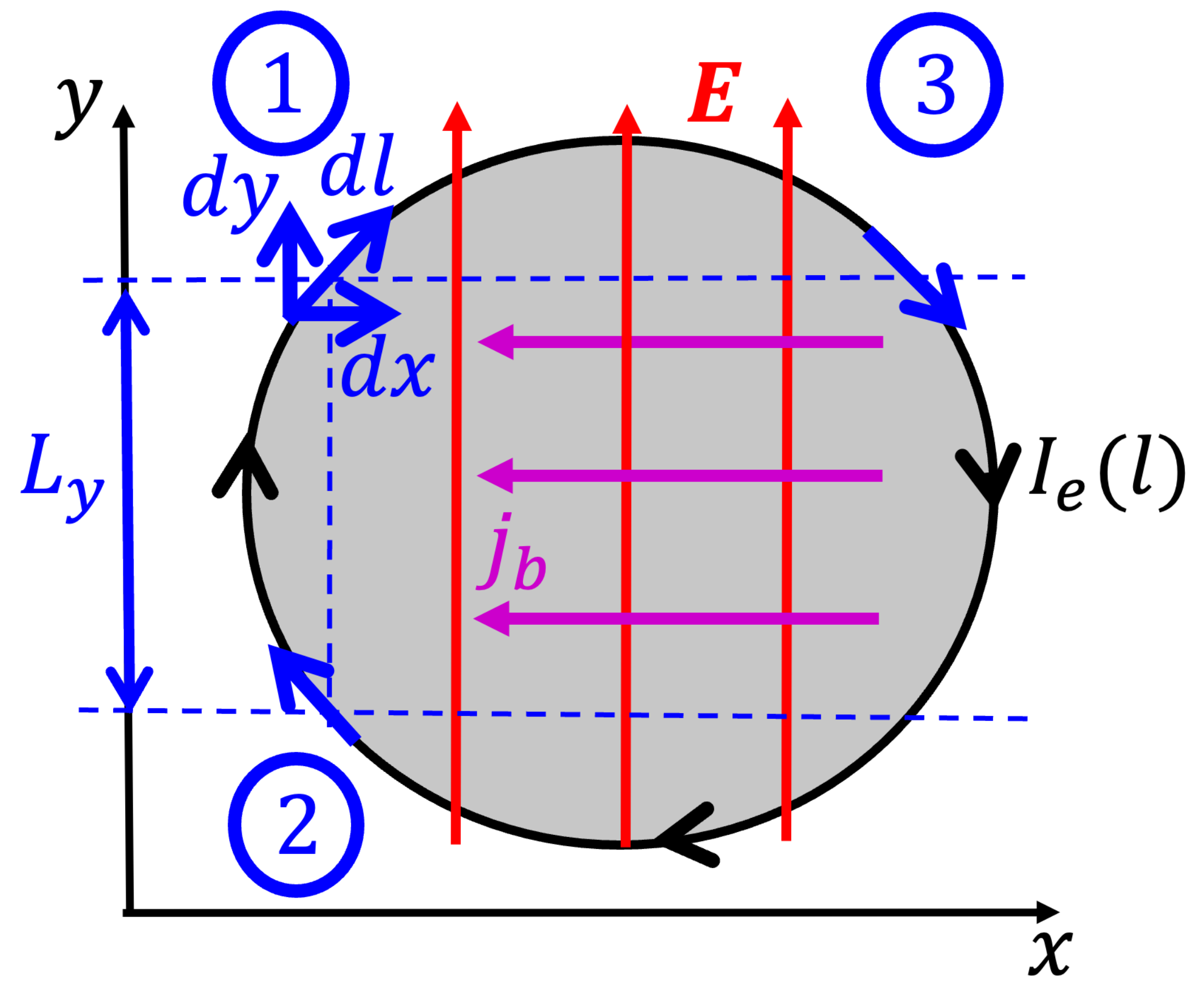}
    \caption{The edge current $I_e(\bm l)$ and the bulk current density $\bm j_b$ in the Chern insulator of a finite size subjected to a uniform electric field $\bm E$.  The black arrows indicate the direction of velocity of the edge modes.} 
	\label{fig:edge}
\end{figure}
%%%%%%%%%%%%%%%%%%%%%%%%%%%%%%%%%%%%%%%%%%%%%%%%%%%%%%%%%%%%%%%%%%%%%%%%%%%%

However, given Eq.~(\ref{sigma_H-dc}), there is a problem that $U_a^{(b)}$ in Eq.~(\ref{Uab_app}) diverges in the dc limit $\omega\to0$, seemingly implying an infinite energy shift at low frequency.  In order to resolve this problem, it is necessary to take into account an additional contribution $U_a^{(e)}$ to the energy shift from the edge currents in a Chern insulator.  Let us consider an isolated sample not connected to any current leads, e.g.,\ a graphene flake, subjected to the spatially uniform electric field $\bm E$ along $y$ axis, as shown in Fig.~\ref{fig:edge}. According to Eqs.~(\ref{Hall-tensor}) and (\ref{sigma_H-dc}), the electric field induces a spatially-uniform bulk Hall current density $\bm j_b$ antiparallel to $x$ axis, which results in the energy shift $U_a^{(b)}$.  However, in an isolated sample in the dc limit, this uniform bulk current has no place to go and must be compensated by an edge current in the opposite direction.  Thus the expectation value $\bm v(t)$ of the velocity operator in Eq.~(\ref{DUAv_app}) must include both the bulk and edge currents
\begin{equation}  \label{v(t)}
  e \bm v(t) = {\mathcal A} \, \bm j_b(t) 
  + \int_{\rm edge} d\bm l \, I_e(\bm l,t).
\end{equation} 
The integral in the last term is taken along the sample edge specified by the coordinate $\bm l$, where $I_e(\bm l,t)$ is the edge current at the position $\bm l$.  Let us use Fig.~\ref{fig:edge} to evaluate this integral in the limit of low frequency.  The Chern insulator has chiral gapless edge modes, whose velocity is indicated by the black arrows in Fig.~\ref{fig:edge}.  The difference between the currents $I_{e}^{(1)}$ and $I_{e}^{(2)}$ carried by the edge modes at points 1 and 2 in Fig.~\ref{fig:edge} is%
\begin{equation}  \label{I_e12}
  I_{e}^{(1)} - I_{e}^{(2)} =  C \, \frac{e^2}{h} \, E_y L_y,
\end{equation} 
where $C$ is the Chern number, and $L_y$ the distance between points 1 and 2 along the direction of the electric field.  The sign in the right-hand side of Eq.~(\ref{I_e12}) is positive, because the electric potential is lower at point 1, thus the occupation of the edge modes is higher than at point 2, so $I_{e}^{(1)}>I_{e}^{(2)}$.  In contrast, the currents carried by the edge modes at points 1 and 3 are equal
\begin{equation}  \label{I_e13}
  I_{e}^{(1)} - I_{e}^{(3)} = 0.
\end{equation} 
Integrating Eq.~(\ref{I_e12}) over $dx$ and Eq.~(\ref{I_e13}) over $dy$, we find that the vector contour integral along the edge in Eq.~(\ref{v(t)}) points along $x$ perpendicularly to $\bm E$ and is proportional to the area ${\mathcal A}$ of the system
\begin{equation}  \label{edge}
  e \bm v_e(t) = \int_{\rm edge} d\bm l \, I_e(\bm l,t) 
  =  {\mathcal A} \, C \, \frac{e^2}{h} \, \left[\bm E(t)\times\hat{\bm z}\right].
\end{equation}
Substituting Eqs.~(\ref{j_b-alpha}), (\ref{Hall-tensor}), and (\ref{edge}) into Eq.~(\ref{v(t)}), we express the expectation value of velocity
\begin{equation}  \label{v_tot}
  e \bm v(t) = {\mathcal A} \, \sigma_H^{(\rm tot)}(\omega)  
  \left[\bm E(t)\times\hat{\bm z}\right],
\end{equation} 
in terms of the effective total ac Hall conductivity $\sigma_H^{(\rm tot)}(\omega)$ that includes both bulk and edge contributions
\begin{align} 
  & \sigma_H^{(\rm tot)}(\omega)
  = \sigma_H^{(b)}(\omega) + \sigma_H^{(e)}(\omega) 
  \label{sigma_two} \\
  & \sigma_H^{(e)}(\omega) =  C \, \frac{e^2}{h} \quad \mbox{at low $\omega$}.
   \label{sigma_H-edge}
\end{align}
Using Eqs.\ (\ref{Xi-Omega-sum}), (\ref{Chern_app}), and (\ref{sigma_H-Xi}) in Eq.~(\ref{sigma_two}), we observe that $\sigma_H^{(\rm tot)}(\omega)$ vanishes at $\omega\to0$ as
\begin{align}  \label{sigma_H-total}
  \sigma_H^{(\rm tot)}(\omega) \approx - \frac{e^2}{\hbar} 
  \sum_{\substack{m \in \left\{\rm emp\right\}\\ n \in \left\{\rm occ\right\}}}
  \int\frac{d^2k}{(2\pi)^2}
  \frac{(\hbar\omega)^2 \, \Xi_{nm}(\bm k)}
  {\varepsilon_{nm}^2(\bm k)-(\hbar\omega+i\delta)^2}.
\end{align}
Then, the full energy shift $U_a=U_a^{(b)}+U_a^{(e)}$ is obtained by replacing $\sigma_H^{(b)}(\omega)\to\sigma_H^{(\rm tot)}(\omega)$ in Eq.~(\ref{Uab_app})
\begin{align}  \label{U-bulk-edge}
  \frac{U_a^{(\rm tot)}}{\mathcal A} 
  = \frac{{\rm Re} \, \sigma_H^{(\rm tot)}(\omega)}{4\omega} \, h_z(\omega).
\end{align}
Using Eq.~(\ref{sigma_H-total}) in Eq.~(\ref{U-bulk-edge}), we observe that the full energy shift vanishes in the dc limit $\omega\to0$ as $U_a^{(\rm tot)}\propto\omega$, in agreement with Eq.~(\ref{energy_shift_time_domain}).  Moreover, using Eq.~(\ref{sigma_H-total}) in Eq.~(\ref{M=sigma_H/omega}), we find that the polarization vector ${\mathcal M}(\omega)\propto\omega$ at low frequency is the same as in Eq.~(\ref{M_app}) for a localized system.  The $\omega^2$ dependence in Eq.~(\ref{sigma_H-total}) can be qualitatively understood as follows.  In the dc limit $\omega=0$, the two contributions in Eq.\ (\ref{sigma_two}) exactly cancel each other, because the total velocity  must vanish in Eq.~(\ref{v_tot}).  However, for a low nonzero frequency, the system develops a nonzero electric dipole moment $\bm d \propto \omega$ perpendicular to $\bm E$ according to Eq.~(\ref{dEM_app}).  This vector $\bm d(t)$ rotates with the frequency $\omega$ of circularly polarized light and has a nonzero time derivative $e\bm v=\dot{\bm d}\propto\omega^2$, which is consistent with Eq.~(\ref{sigma_H-total}) at low $\omega$.

The quantized formula (\ref{sigma_H-edge}) for the edge contribution is applicable only at low frequency and is expected to be suppressed at high frequency.  A detailed quantitative study of the frequency dependence of the edge contribution is beyond the scope of this paper.

%%%%%%%%%%%%%%%%%%%%%%%%%%%%%%%%%%%%%%%%%%%%%%%%%%%%%%%%%%%%%%%%%%%%%%%%%%%%%
\section{Valley-selective optics on the honeycomb lattice} 
\label{sec:valley}
%%%%%%%%%%%%%%%%%%%%%%%%%%%%%%%%%%%%%%%%%%%%%%%%%%%%%%%%%%%%%%%%%%%%%%%%%%%%%

For applications to graphene and other 2D materials, let us focus on the total Stark energy shift without separating it into the symmetric and antisymmetric parts.  In 2D, Eqs.~(\ref{chi0_app}), (\ref{DE_n}), (\ref{UV0_app}), and (\ref{DE-intra-3D}) give the energy shift of the state $|n,\bm k\rangle$
\begin{widetext}
\begin{align}  
  & \Delta \varepsilon_n(\bm k) = -\frac{e^2\Omega_n(\bm k)}{4\,\omega} \epsilon^{\alpha\beta} \,{\rm Im}\left[ E_\alpha^\ast(\omega)E_\beta(\omega)\right]  - \frac{e^2 }{4}\,{\rm Re}\sum_{m\neq n}\left[ \frac{r^{\alpha}_{nm}(\bm k)r^\beta_{mn}(\bm k)}{\varepsilon_{mn}(\bm k)-\omega}
  + \frac{\left[r^{\alpha}_{nm}(\bm k)r^\beta_{mn}(\bm k)\right]^\ast}{\varepsilon_{mn}(\bm k)+\omega} \right]E_\alpha^\ast(\omega)E_\beta(\omega),
\label{full_stark_shift_2D}
\end{align}
\end{widetext}
where $\delta$ is omitted in the denominators.  The second term in Eq.~(\ref{full_stark_shift_2D}) is the interband contribution from the conventional second-order perturbation theory. The terms with $\left[\varepsilon_{mn}(\bm k)-\omega\right]$ and $\left[\varepsilon_{mn}(\bm k)+\omega\right]$ in the denominators are well known as the Stark \cite{Sie2015} and the Bloch-Siegert \cite{Sie2017} contributions, respectively. In contrast, the first term represents the intraband contribution to the energy shift due to the Berry curvature derived in Eqs.~(\ref{DE-intra-3D}) and (\ref{m'=m}), which does not seem to be well recognized in the literature.  This term is absent for linearly polarized light, and its sign depends on relative orientation of the helicity and Berry curvature vectors.

Let us apply this general result to electrons on the honeycomb lattice in the presence of a staggered gap $\pm\Delta$.  The low-energy Hamiltonian in the sublattice basis corresponds to massive Dirac electrons
\begin{align}  \label{ham_gr_BN}
    H = \left(\begin{array}{cc}
         \Delta & v(sk_y+ik_x) \\
         v(sk_y-ik_x) & -\Delta 
    \end{array}\right).
\end{align}
Here the origin of momentum $\bm k =(k_x,k_y)$ is chosen at the $K$ and $K'$ points in the Brillouin zone labeled by the index $s=\pm1$.  The Dirac velocity $v = 3aw_1/2$ is related to the nearest-neighbor distance $a$ and the hopping amplitude $w_1$.  The off-diagonal elements at the $K$ and $K'$ points have opposite phase windings, defined as the contour integral $\oint d\bm k\cdot \partial_{\bm k} \, {\rm Arg}[v(sk_y-ik_x)] = 2\pi s$ in the counterclockwise direction.  This equation can be taken as the definition of the $K$ and $K'$ points.  From Hamiltonian (\ref{ham_gr_BN}), we find the wavefunctions for the valence  and conduction bands at $k \ll \Delta/v$
\begin{equation*}
    | v \rangle \approx \left(\begin{array}{c}
         \frac{v}{2\Delta} (-sk_y-ik_x) \\
         1 
    \end{array}\right), \,\,
        | c \rangle \approx \left(\begin{array}{c}
         1 \\
         \frac{v}{2\Delta} (sk_y-ik_x) 
    \end{array}\right), 
\end{equation*}
and obtain the matrix elements \cite{WangYao2008} $r_{cv}^\alpha = i{\langle c|\partial^\alpha|v\rangle}$ as $x_{cv}=is y_{cv}=r_0= v/2\Delta$. 
They allow us to evaluate the product of matrix elements appearing in Eq.~(\ref{full_stark_shift_2D}) 
\begin{equation}  \label{matr_el_prod}
   r_{vc}^\alpha r_{cv}^\beta = \frac{v^2}{4\Delta^2}\,\left(\delta^{\alpha\beta}-i\,s\,\epsilon^{\alpha\beta}\right),
\end{equation}
as well as the Berry curvature of the valence band
\begin{equation}
   \Omega_v = -\epsilon_{\alpha\beta}\,{\rm Im}[r_{vc}^\alpha r_{cv}^\beta] = \frac{s\,v^2}{2\Delta^2}
   \label{Ber_curv_2band}
\end{equation}
at the $K$ and $K'$ points.  Remarkably, the symmetric and antisymmetric parts in Eq.~(\ref{matr_el_prod}) are equal in magnitude.  Substituting Eqs.~(\ref{matr_el_prod}) and (\ref{Ber_curv_2band}) in Eq.~(\ref{full_stark_shift_2D}) and using the circularly polarized electric field $E_{\pm}(\omega)$ from Eq.~(\ref{E+-}), we obtain the energy shift of the valence band at the $K$ and $K'$ points
\begin{align}
 \Delta\varepsilon_v =& \frac{\hbar^2 e^2v^2 E^2_{\pm}(\omega)}{4\Delta^2} \left[\mp \frac{s}{\hbar\omega} - \frac{1}{2\Delta \mp s\, \hbar\omega}\right]\label{stark_2band} \\
 =& \mp s\frac{\hbar \,e^2v^2  E_{\pm}^2(\omega)}{2\,\omega\Delta\left(2\Delta \mp s\,\hbar\omega\right)}, \label{stark_2band_b} 
\end{align}
where we restored $\hbar$ in the expression. The second term in the brackets in Eq.~(\ref{stark_2band}) represents either the Stark or the Bloch-Siegert contribution from Eq.~(\ref{full_stark_shift_2D}) depending on the valley index $s=\pm1$.  The fact that each contribution is present only for one of the two valleys is a manifestation of the well-known valley selectivity of coupling to circularly polarized light \cite{Kim2014,Sie2015,Sie2017,Mak2018}.  As usual for the second-order perturbation theory at a subgap frequency $\hbar\omega < 2\Delta$, each contribution produces a negative shift of the valence band energy $\Delta \varepsilon_v<0$ and a positive shift for the conduction band $\Delta \varepsilon_c>0$, thus resulting in repulsion of energy levels and an increase in the energy gap for both valleys.

In contrast, the first term in Eq.~(\ref{stark_2band}), originating from the Berry curvature contribution in Eq.~(\ref{full_stark_shift_2D}), has opposite signs for two valleys.  Moreover, this term $\propto 1/\hbar\omega$ is greater in magnitude than the Bloch-Siegert term $\propto 1/(\hbar\omega+2\Delta)$.  Thus, for a subgap frequency $\hbar\omega < 2\Delta$, the sign of the energy shift~(\ref{stark_2band_b}) is opposite for two valleys, and the energy gap \textit{decreases} for the valley coupling to the Bloch-Siegert term.

The valley-selective optical Stark and Bloch-Siegert energy shifts were measured experimentally in WS$_2$ and WSe$_2$ monolayers \cite{Kim2014,Sie2015,Sie2017}.  Upon illumination by a circularly polarized pump pulse with a subgap frequency $\hbar\omega<2\Delta$, an \textit{increase} in exciton energy was observed for both valleys by using probe pulses of either circular polarization.  A good quantitative fit was obtained by using only the Stark or the Bloch-Siegert term in Eq.~(\ref{stark_2band}) for each valley \cite{Sie2017}, without recognition of the first term due to the Berry curvature.

We believe that the Berry curvature term $\propto 1/\omega$ in Eq.~(\ref{stark_2band}) was not observed in experiments \cite{Kim2014,Sie2015,Sie2017}, because they measured the energy shift of an exciton, i.e.,\ a bound state of an electron-hole pair, rather than the band edge of light absorption, corresponding to generation of free, unbound electron-hole pairs.  The free electron and hole have anomalous velocities of opposite signs in Eq.~(\ref{v-d-nk}), i.e.,\ they move away from each other, because of the opposite signs of the Berry curvature for the valence and conduction bands.  In the ground $s$-wave state of an exciton, the electron and hole are bound together by Coulomb attraction and do not move relative to each other, so they loose their anomalous velocities.  In other words, an exciton is an electrically neutral composite object, so its net Berry curvature is zero.  Once the anomalous velocity is lost due to excitonic binding, the corresponding contribution to the energy shift obtained by time averaging in Eq.~(\ref{DUAv_app}) is also lost.  In contrast, the valley Hall effect, observed in $\rm MoS_2$ \cite{Mak2014,Lee2016} upon generation of free carriers in one valley by circularly polarized light, is an experimental manifestation of the valley-dependent anomalous velocity due to the Berry curvature.  But the valley Hall effect is only possible for free electrons and holes, not for bound excitons.  On the other hand, the Berry curvature does affect the energies of excited states of an exciton with a nonzero angular momentum, where the electron and hole move relative to each other \cite{Srivastava2015,Zhou2015}.

\begin{figure} 
\includegraphics[width=\linewidth]{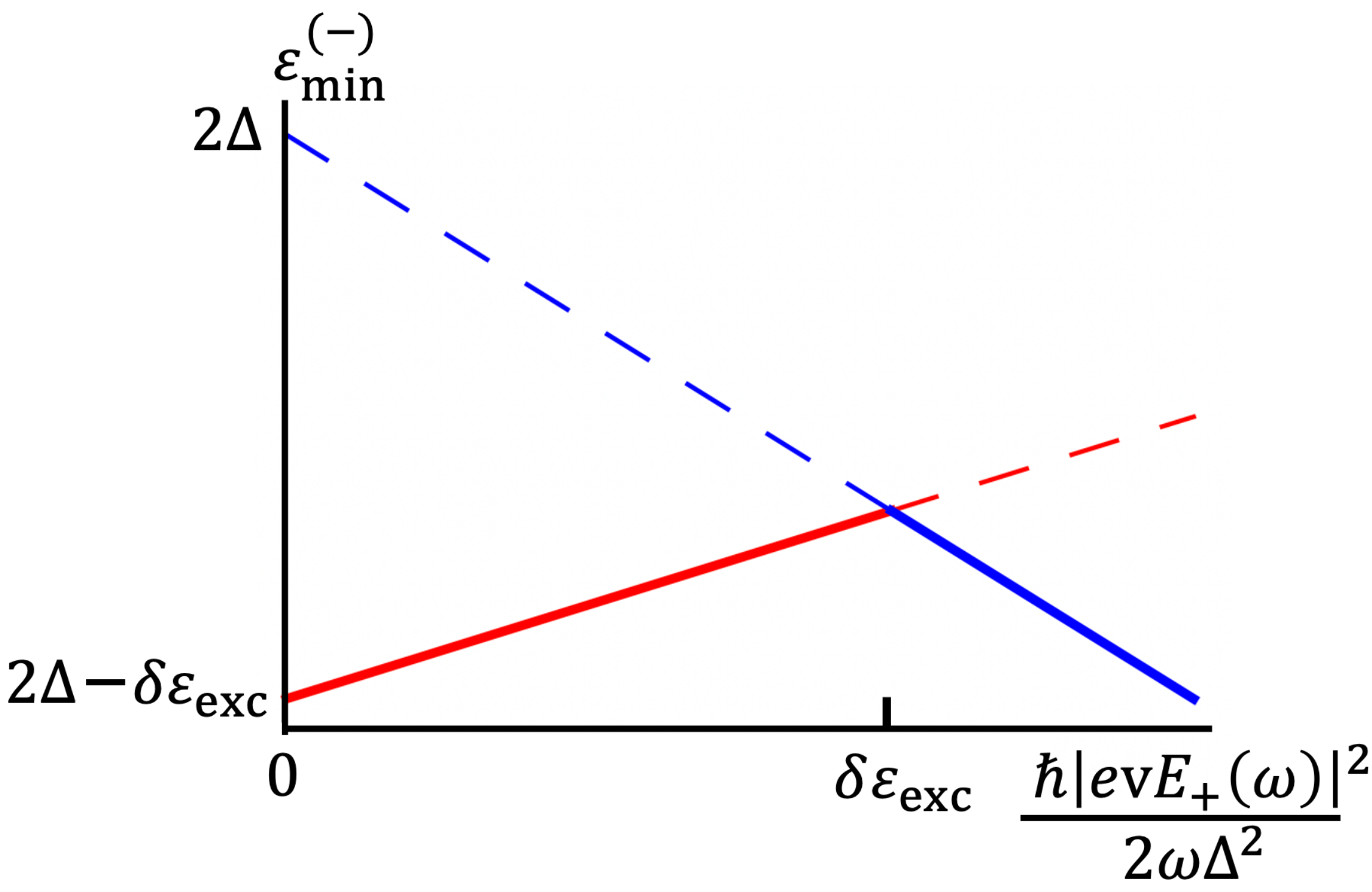}
\caption{The minimal excitation energy $\varepsilon_{\rm min}^{(-)}$, as measured by a circularly polarized probe, vs the pump fluence $E_+^2$ of the opposite circular polarization.  The red line of positive slope represents the top entry in Eq. (\ref{piece-wise}), corresponding to exciton formation.  The blue line of negative slope represents the bottom entry in Eq. (\ref{piece-wise}), corresponding to unbound electrons and holes.  At the intersection of the two lines, the threshold condition (\ref{Berry>exciton}) for spontaneous dissociation of excitons is satisfied.  The parameters chosen for this graph are $\hbar\omega = \Delta$ and $\delta\varepsilon_{\rm exc} = 0.2 \Delta$.}
\label{fig:emin}
\end{figure}

Although the decrease in the energy gap in one valley predicted by Eq.~(\ref{stark_2band_b})  has not been observed experimentally so far, it may be detected for a sufficiently high fluence of the circularly polarized pump pulse.  Formation of an exciton is energetically favorable because of the energy decrease $\delta\varepsilon_{\rm exc}$ due to Coulomb attraction.  On the other hand, the Berry curvature term in Eq.~(\ref{stark_2band}) also results in energy decrease for one valley, but only for unbound electrons and holes.  Thus, there is competition between the two mechanisms for decreasing energy.  The energy decrease due to the Berry mechanism is greater when the fluence is sufficiently high and the following condition is satisfied
\begin{align}  \label{Berry>exciton} 
 \frac{\hbar\, e^2v^2 E^2_{\pm}(\omega)}{2\Delta^2\omega} > \delta\varepsilon_{\rm exc}.
\end{align}
When Eq.~(\ref{Berry>exciton}) is satisfied, excitons in the valley coupling to the Bloch-Siegert term will spontaneous dissociate into free electrons and holes, because the dissociated state has lower energy than a bound state.  The minimal energy $\varepsilon_{\rm min}$ needed to create an excitation in one of the two valleys is given by the piecewise continuous formula
\begin{align}  \label{piece-wise}
  \varepsilon_{\rm min}^{(-)} = 2\Delta 
  + \frac{\hbar^2\,e^2v^2 E^2_+(\omega)}{2\Delta^2(2\Delta+\hbar\omega)} 
  - \left\{ {\delta\varepsilon_{\rm exc} \atop 
  \hbar|evE_+(\omega)|^2 / 2\Delta^2\omega } \right. , 
\end{align}
where, for concreteness, we took the light polarization $+$ and the valley $s=-1$.  The top line in Eq.~(\ref{piece-wise}) is applicable for fluence below the threshold and the bottom line above the threshold given by Eq.~(\ref{Berry>exciton}).  The first two terms in Eq.~(\ref{piece-wise}) are the bare band gap and the Bloch-Siegert shift. A graph of the minimal excitation energy $\varepsilon_{\rm min}^{(-)}$ vs the fluence $E_+^2(\omega)$ is shown in Fig. \ref{fig:emin}.  To verify Eq.~(\ref{piece-wise}) experimentally, a probe of the opposite circular polarization to the pump should be used.  However, it may be difficult to achieve the threshold condition (\ref{Berry>exciton}) experimentally.  An optically induced energy shift $\sim$15 meV was demonstrated in Ref.~\cite{Sie2017}, whereas the exciton binding energy is in the range 0.3--1 eV \cite{Mak2018}.

%%%%%%%%%%%%%%%%%%%%%%%%%%%%%%%%%%%%%%%%%%%%%%%%%%%%%%%%%%%%%%%%%%%%%%%%%%%%%
%%%%%%%%%%%%%%%%%%%%%%%%%%%%%%%%%%%%%%%%%%%%%%%%%%%%%%%%%%%%%%%%%%%%%%%%%%%%

%%%%%%%%%%%%%%%%%%%%%%%%%%%%%%%%%%%%%%%%%%%%%%%%%%%%%%%%%%%%%%%%%%%%%%%%%%%%%
\section{Application to twisted graphene multilayers} 
\label{sec:graphene}
%%%%%%%%%%%%%%%%%%%%%%%%%%%%%%%%%%%%%%%%%%%%%%%%%%%%%%%%%%%%%%%%%%%%%%%%%%%%%

Now let us discuss how this framework can be applied to control spontaneous valley polarization and concomitant orbital magnetization for a twisted graphene bilayer on the boron nitride (BN) substrate at the filling factor $\nu=3$.  We follow the interpretation presented in Ref.~\cite{Serlin2020} on the basis of theoretical models \cite{Song2015,Zhang2019PRR,Bultnick2020}.  The staggered sublattice potential induced by the BN substrate separates the valence and conduction bands of graphene by the energy gap $2\Delta$ and generates a nonzero Berry curvature at the $K$ and $K'$ points.  The periodic potential produced by the moir\'e superlattice folds the valence and conduction bands into mini-bands at the $K$ and $K'$ valleys, where the lowest mini-bands acquire the Chern number $C=\pm 1$.  At the filling factor $\nu=4$ per moir{\'e} unit cell, the four mini-bands in the conduction band are fully occupied (for valleys $K$ and $K'$, and spins up and down), so the system is a trivial band insulator.  When the mini-bands above and below the energy gap are occupied, the system does not experience an energy shift in circularly polarized light.

In contrast, at $\nu = 3$, Coulomb interaction between electrons induces spontaneous valley polarization.  In this case, the two degenerate mini-bands (for spins up and down) produced by the moir\'e superlattice in the conduction band of, e.g.,\ the $K'$ valley remain fully occupied.   But in the $K$ valley, one of the two mini-bands (with a given spin orientation) in the conduction band is pushed above the Fermi level due to Coulomb repulsion between electrons via the Hund or Stoner mechanisms, so this mini-band becomes empty (see Fig.~1(c) in Ref.~\cite{Serlin2020}).  Thus the system develops a spontaneous valley polarization due to electron correlations.  This empty mini-band and the corresponding filled mini-band in the valence band, treated within a two-band model, do experience the ac Stark energy shift $U_a$ in Eq.~(\ref{stark_two_bands}) due to the Berry curvature of the $K$ valley.  By properly choosing the sign of helicity of circularly polarized light, it is possible to raise the energy $U_a$ of this valley configuration and make it energetically unfavorable, thus forcing a spontaneous switch to the opposite valley polarization.

This result can be also interpreted in terms of valley-selective optics discussed in Sec.~\ref{sec:valley}.  Let us consider the case of high frequency $\hbar\omega\gg|\varepsilon_{cv}(\bm k)|$, given that the energy gap $\Delta\sim 1$~meV is quite small here.  In this limit, Eq.~(\ref{stark_2band_b}) becomes
\begin{align}
    \Delta\varepsilon_v \approx \frac{e^2\, v^2  |E_{\pm}(\omega)|^2}{2\Delta} \left[\frac{1}{\omega^2}\pm \frac{2s\Delta}{\hbar\omega^3}\right]. \label{valence_band_high_freq}
\end{align}
The second term in Eq.~(\ref{valence_band_high_freq}) shows that the energy shift is different for the two valleys and depends on helicity of circularly polarized light. Thus applying the right $E_{+}(\omega)$ or left $E_{-}(\omega)$ circularly polarized light renders the $K'$ ($s=-1$) or $K$ ($s=1$) valley more energetically favorable, which would trigger a spontaneous switch of valley polarization.  This is our proposed mechanism for optical control of valley polarization and orbital magnetization in graphene multilayers using circularly polarized light on the basis of the ac Stark effect.

Let us crudely estimate the magnitude of the proposed effect using  Eq.~(\ref{stark_two_bands}).  We estimate the antisymmetric ac Stark energy shift per unit cell as $\tilde U^{(1)}_a \sim (eEa_m)^2C/4\pi\Delta$ for moderate frequencies $\hbar\omega \sim \Delta$ and $\tilde U^{(2)}_a \sim (\Delta/\hbar\omega)^3 \,\tilde U^{(1)}_a $ in the high-frequency limit $\hbar \omega \gg \Delta$.  Here $a_m$, $\Delta$ and $C$ are the moir\'e unit cell size, the energy gap, and the Chern number, respectively.  Taking the characteristic values $\Delta \sim 1$~meV, $a_m \sim 10$~nm, $C=1$, and $E = 3\times 10^4$~V/m (for a $0.01$~W laser with the spot size $d=0.1$~mm), we find the Stark energy $\tilde U_a^{(1)} \sim  10^{-5}$~eV at frequency $\omega = \Delta\sim 1\, {\rm meV} = 0.24\, {\rm THz}$.  For a laser of visible light $\hbar\omega \sim 2$ eV, this estimate is reduced by a factor of $(\Delta/\hbar\omega)^3 \sim  \left(5\times 10^{-4}\right)^3 \sim 10^{-10}$, thus producing $\tilde U_a^{(2)} \sim 10^{-15}$~eV.  In comparison, the typical energy shift due to a magnetic field is $\mu_B B \sim 5.8\times 10^{-5}\,{\rm eV}\times B~[\rm T]$ per unit cell carrying a magnetic moment of the order of the Bohr magneton \cite{Polshyn2020}. Thus, the estimate $\tilde U^{(1)}_{a}$ corresponds to the effective magnetic field $B^{(1)} \sim 0.1$~T.  Experimentally~\cite{Polshyn2020}, magnetization was switched by a magnetic field as low as $B=0.08$~T. So, the effect of circularly polarized light is comparable to the magnetic fields used in experiments \cite{Serlin2020,Polshyn2020} to control magnetization.  With a much higher pulsed laser power, an effective magnetic field $\sim50$~T has been achieved in the optical Stark effect in transition metal dichalcogenides \cite{Kim2014,Sie2015,Sie2017}. However, as discussed in Sec.~\ref{sec:experiment}, the most realistic experimental protocol involves cooling the system through the second-order Landau phase transition, rather than trying to switch magnetization at low temperature.  At the transition temperature, even a small time-reversal-symmetry-breaking Stark shift is sufficient to force the system to choose a desired sign of magnetization.

%%%%%%%%%%%%%%%%%%%%%%%%%%%%%%%%%%%%%%%%%%%%%%%%%%%%%%%%%%%%%%%%%%%%%%%%%%%%%
\section{Conclusions}
\label{sec:conclusions}
%%%%%%%%%%%%%%%%%%%%%%%%%%%%%%%%%%%%%%%%%%%%%%%%%%%%%%%%%%%%%%%%%%%%%%%%%%%%%

For the systems that break time-reversal symmetry, the energy of interaction with an external ac electric field $\bm E(\omega)$ contains an antisymmetric term $U_a$ in Eq.~(\ref{energy_shift}) that couples to the helicity (\ref{helicity}) of circularly polarized light.  Taking advantage of this interaction energy $U_a$, we propose an experimental protocol for switching the orbital magnetization of a Chern insulator using circularly polarized light, as discussed in Sec.~\ref{sec:experiment}.  A gradual lowering of the sample temperature from $T>T_c$ to $T<T_c$ in the presence of circularly polarized light forces the system to select a particular sign of the orbital magnetization and the spontaneous Hall resistance $R_H$ at the transition, depending on the helicity of light.  Alternatively, applying circularly polarized light of the appropriate helicity at a low temperature, it is possible to raise the energy of a given spontaneous valley polarization and make it energetically unfavorable, thus forcing a switch to the opposite valley polarization.  Both mechanisms realize optical control of topological memory based on spontaneous orbital magnetization.  Moreover, by proper positioning of two laser beams with opposite helicities, it would be possible to create a chiral domain wall between domains with opposite orbital magnetizations at a predetermined location.  There is great interest in using the topologically protected one-way electronic edge states localized on such domain walls for potential applications.

We derived explicit microscopic expressions for the antisymmetric ac Stark energy shift of single electron states and the total energy of the system.  
In contrast to the localized systems, such as atoms, discussed in Sec.~\ref{sec:discrete}, the energy shift for a Chern insulator contains an additional intraband contribution (\ref{DE-intra-3D}) originating from the Berry curvature and related to the anomalous velocity.  This term is important for reconciling two derivations of the ac Stark energy shift, based on the dynamical polarizability and the ac Hall conductivity, connected by Eq.~(\ref{M=sigma_H/omega}).  The vector $\bm{\mathcal M}_b(\omega)$ in Eq.~(\ref{energy_shift}) is expressed as an integral over the Brillouin zone involving either the Berry connections in Eq.~(\ref{M-2D}) in general case or the Berry curvature in Eq.~(\ref{M_Berry_app}) for a two-band Chern insulator.  Moreover, we predict spontaneous dissociation of excitons in one valley of a gapped honeycomb lattice for a sufficiently high fluence (\ref{Berry>exciton}) of circularly polarized light, due to the intraband contribution to the energy shift.

{\it Note Added in Proofs.} There is experimental evidence for
orbital magnetic order also in the kagome material KV$_3$Sb$_5$
\cite{Jiang2021,Mielke2022}. Our proposal for optical control of orbital magnetization is applicable to this material as well.

\begin{acknowledgments}
We thank G.~Polshyn for stimulating discussions that motivated this work. SP was supported by the U.S.\ Department of Energy (DOE), Office of Science, Basic Energy Sciences (BES) under Award No. DE-SC0020221.
\end{acknowledgments}

%%%%%%%%%%%%%%%%%%%%%%%%%%%%%%%%%%%%%%%%%%%%%%%%%%%%%%%%%%%%%%%%%%%%%%%%%%%%%
\appendix
%%%%%%%%%%%%%%%%%%%%%%%%%%%%%%%%%%%%%%%%%%%%%%%%%%%%%%%%%%%%%%%%%%%%%%%%%%%%%

%%%%%%%%%%%%%%%%%%%%%%%%%%%%%%%%%%%%%%%%%%%%%%
\section{An alternative derivation of the ac Stark effect}
\label{sec:Alternative}
%%%%%%%%%%%%%%%%%%%%%%%%%%%%%%%%%%%%%%%%%%%%%%

An alternative derivation of the ac Stark energy shift, also common in the literature \cite{SakuraiBook,Kobe1983,haas2006}, can be obtained by calculating the second-order correction to the wavefunction amplitude $c_n(t)$ of the ground state.  Substituting Eq.~(\ref{c_m_app}) into Eq.~(\ref{schrod_c}), we find \begin{equation}  \label{c._app}
\begin{aligned}  
    i\dot c_n(t) \approx & - \sum_{m\neq n} \left[
    \frac{V^\dagger_{nm}V_{mn}}{\varepsilon_{mn}-\omega-i\delta}
    + \frac{V_{nm}V^\dagger_{mn}}{\varepsilon_{mn}+\omega-i\delta}\right. \\ 
    & \left. + \frac{V_{nm}V_{mn} \, e^{-2i\omega t}}
    {\varepsilon_{mn}-\omega-i\delta} 
    + \frac{V^\dagger_{nm}V^\dagger_{mn} \, e^{2i\omega t}}
    {\varepsilon_{mn}+\omega-i\delta}\right] e^{2t\delta} . 
\end{aligned}
\end{equation}
The fast-oscillating terms at the frequency $2\omega$ drop out upon averaging over time.  Separating the real and imaginary parts of the remaining terms in the right-hand side of Eq.~(\ref{c._app}) and setting $\delta\to 0$ in the real part, we find
\begin{equation} \label{c.ReIm_app}
    i\dot c_n(t) \approx \Delta\varepsilon_n 
    -i\delta e^{2t\delta} \sum_{m\neq n} \left[
    \frac{|V_{mn}|^2}{|\varepsilon_{mn}-\omega|^2}
    + \frac{|V_{mn}^\dag|^2}{|\varepsilon_{mn}+\omega|^2}
      \right],
\end{equation}
where we assume that $\omega\neq|\varepsilon_{mn}|$.  The first term
\begin{equation}  \label{e_renorm}
   \Delta\varepsilon_n = - \sum_{m\neq n}\left[
   \frac{|V_{mn}|^2}{\varepsilon_{mn}-\omega}
   + \frac{|V_{mn}^\dag|^2}{\varepsilon_{mn}+\omega}\right]
\end{equation}
in the energy shift of the state $n$.  Indeed, the amplitude of this state should be written as
\begin{equation}  \label{c+tilde_app}
c_n(t)=\tilde c_n(t)\,e^{-it\Delta\varepsilon_n}.  
\end{equation}
Substituting Eq.~(\ref{c+tilde_app}) into Eq.~(\ref{schrod_c}) for $\dot c_n(t)$ produces the term $\Delta\varepsilon_n$ in the left-hand side, which cancels $\Delta\varepsilon_n$ in the right-hand side of Eq.~(\ref{c.ReIm_app}), whereas the exponential factor produces a small renormalization of $\varepsilon_{mn}$ in Eq.~(\ref{c_m_app}).  Equation (\ref{e_renorm}) for the energy shift agrees with Eq.~(\ref{UV0_app}).

The rest of Eq.~(\ref{c.ReIm_app}) generates an equation for $\tilde c_n(t)$
\begin{equation} \label{c.tilde_app}
    \frac{d\tilde c_n(t)}{dt} \approx 
    - \delta e^{2t\delta} \sum_{m\neq n} \left[
    \frac{|V_{mn}|^2}{|\varepsilon_{mn}-\omega|^2}
    + \frac{|V_{mn}^\dag|^2}{|\varepsilon_{mn}+\omega|^2} \right].
\end{equation}
Integrating Eq.~(\ref{c.tilde_app}) over time and then setting $\delta\to 0$, we find
\begin{align} 
    & \tilde c_n \approx 1 - \frac12 \sum_{m\neq n} \left[
    \frac{|V_{mn}|^2}{|\varepsilon_{mn}-\omega|^2}
    + \frac{|V_{mn}^\dag|^2}{|\varepsilon_{mn}+\omega|^2} \right],
    \label{c-tilde_app} \\
    & |c_n|^2 \approx 1 - \sum_{m\neq n} \left( |c_{m-}|^2+|c_{m+}|^2 \right).
    \label{c2-tilde_app}
\end{align}
Equation (\ref{c2-tilde_app}) shows that the decrease in the probability $|c_n|^2$ is equal to the sum of probabilities $|c_{m\mp}|^2$ in Eq.~(\ref{c_m_app}) for the other states, as mentioned above Eq.~(\ref{U1_app}).

For the states satisfying the resonant condition $\omega=|\varepsilon_{mn}|$, the imaginary part of Eq.~(\ref{c._app}) represents the depopulation rate of the initial state due to real (as opposed to virtual) transitions to other states with the absorption or emission of a photon \cite{SakuraiBook}.  The difference in the signs of $i\delta$ in the denominators of Eq.~(\ref{UV0_app}) and the time-averaged Eq.~(\ref{c._app}) can be understood as follows.  The imaginary part of the polarizability tensor (\ref{chi0_app}) appearing in Eq.~(\ref{energy_shift_def}) represents the rate of energy absorption, where absorption and emission of photons is counted with opposite signs \cite{LandauVol5}.  In contrast, the imaginary part of Eq.~(\ref{c._app}) represents the depopulation rate of the initial state, where absorption and emission of photons is counted with the same sign.  When the system is in the ground state at zero temperature, Eqs.~(\ref{energy_shift_def}) and (\ref{c._app}) are consistent, because only absorption of photons is possible, not emission.  But in general, it is preferable to use the polarizability tensor, which satisfies the Kramers-Kronig relation and whose imaginary part has clear physical interpretation in terms of the energy absorption rate.

%%%%%%%%%%%%%%%%%%%%%%%%%%%%%%%%%%%%%%%%%%%%%%%%%%%%%%%%%%%%%%%%%%%%%%%%%%%%%
\begin{figure}
	(a)\includegraphics[width=0.8\linewidth]{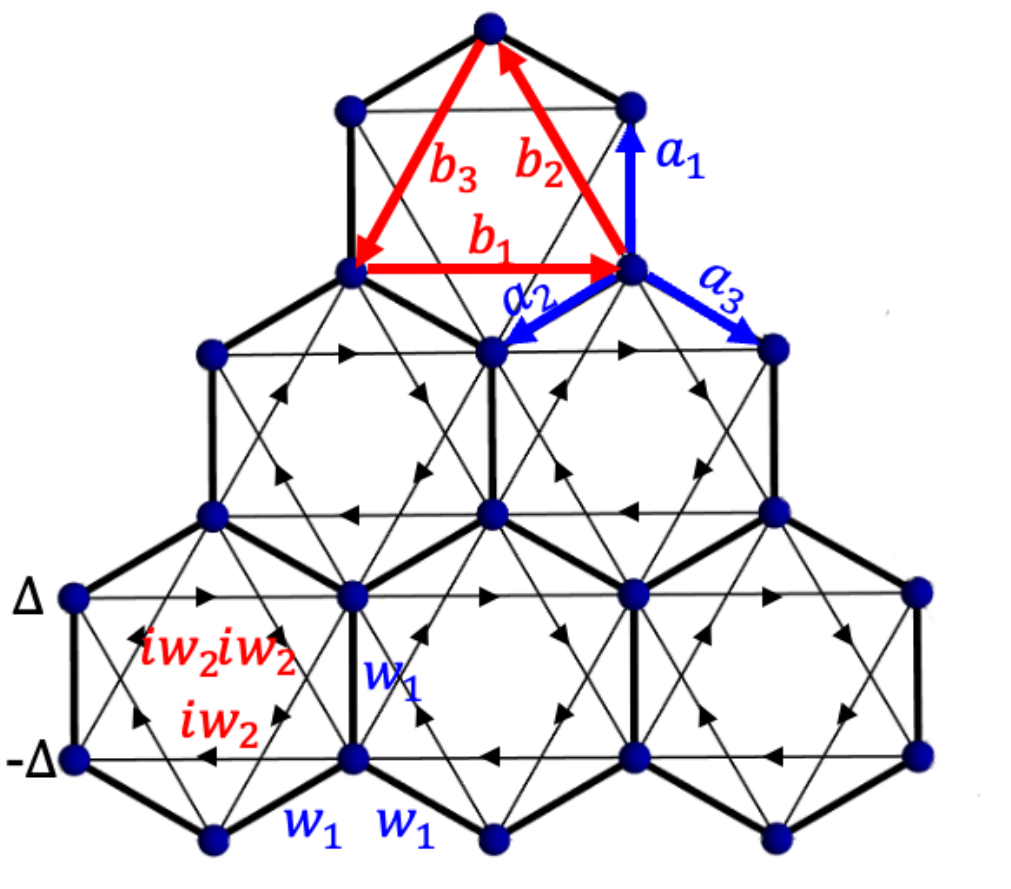} \\
	(b)\includegraphics[width=0.8\linewidth]{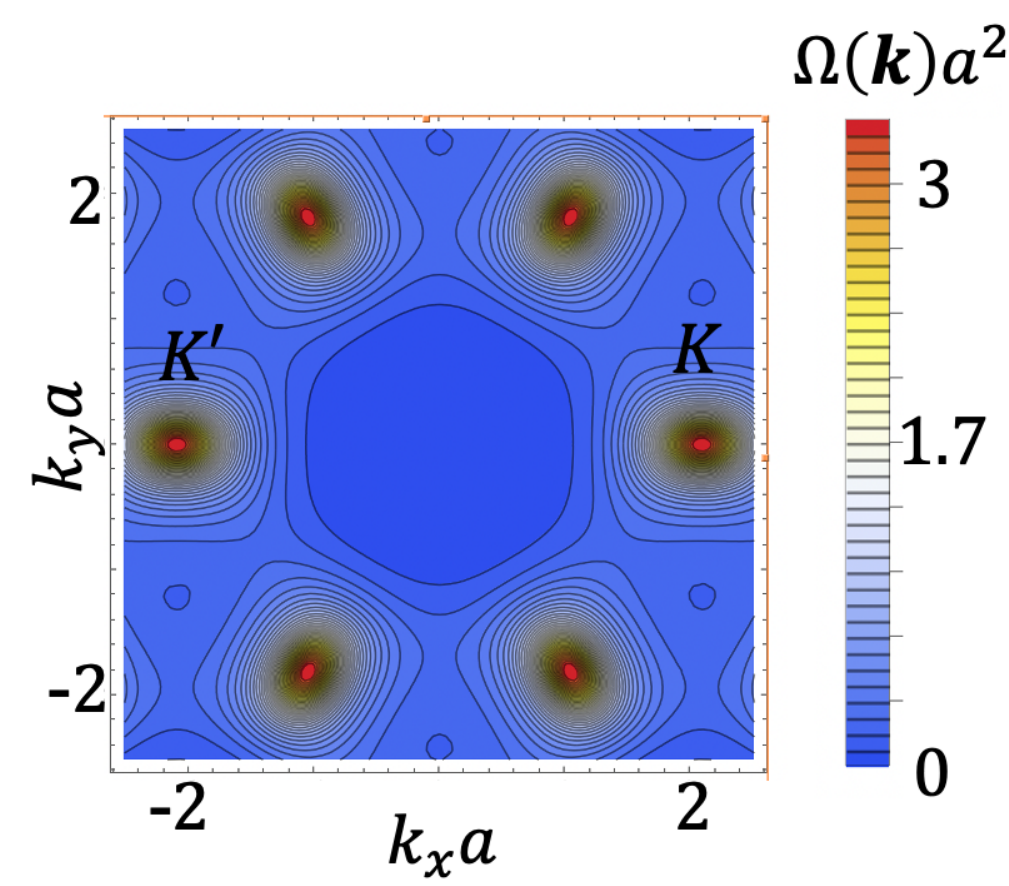} 
	\caption{(a) Haldane model on the honeycomb lattice. 
	(b) Berry curvature $\Omega_{v}(\bm k)$ of the valence band vs momentum $\bm k$ in the topologically nontrivial case $C=1$. The chosen parameters are $w_2 = w_1/2$ and $\Delta = 0$.} 
	\label{fig:lattice_and_spectrum}
\end{figure}
%%%%%%%%%%%%%%%%%%%%%%%%%%%%%%%%%%%%%%%%%%%%%%%%%%%%%%%%%%%%%%%%%%%%%%%%%%%%%
%%%%%%%%%%%%%%%%%%%%%%%%%%%%%%%%%%%%%%%%%%%%%%%%%%%%%%%%%%%%%%%%%%%%%%%%%%%%%
\section{Application to the Haldane model} 
\label{appendix:Haldane}
%%%%%%%%%%%%%%%%%%%%%%%%%%%%%%%%%%%%%%%%%%%%%%%%%%%%%%%%%%%%%%%%%%%%%%%%%%%%%
%%%%%%%%%%%%%%%%%%%%%%%%%%%%%%%%%%%%%%%%%%%%%%%%%%%%%%%%%%%%%%%%%%%%%%%%%%%%%
\begin{figure}
	(a)\includegraphics[width=0.9\linewidth]{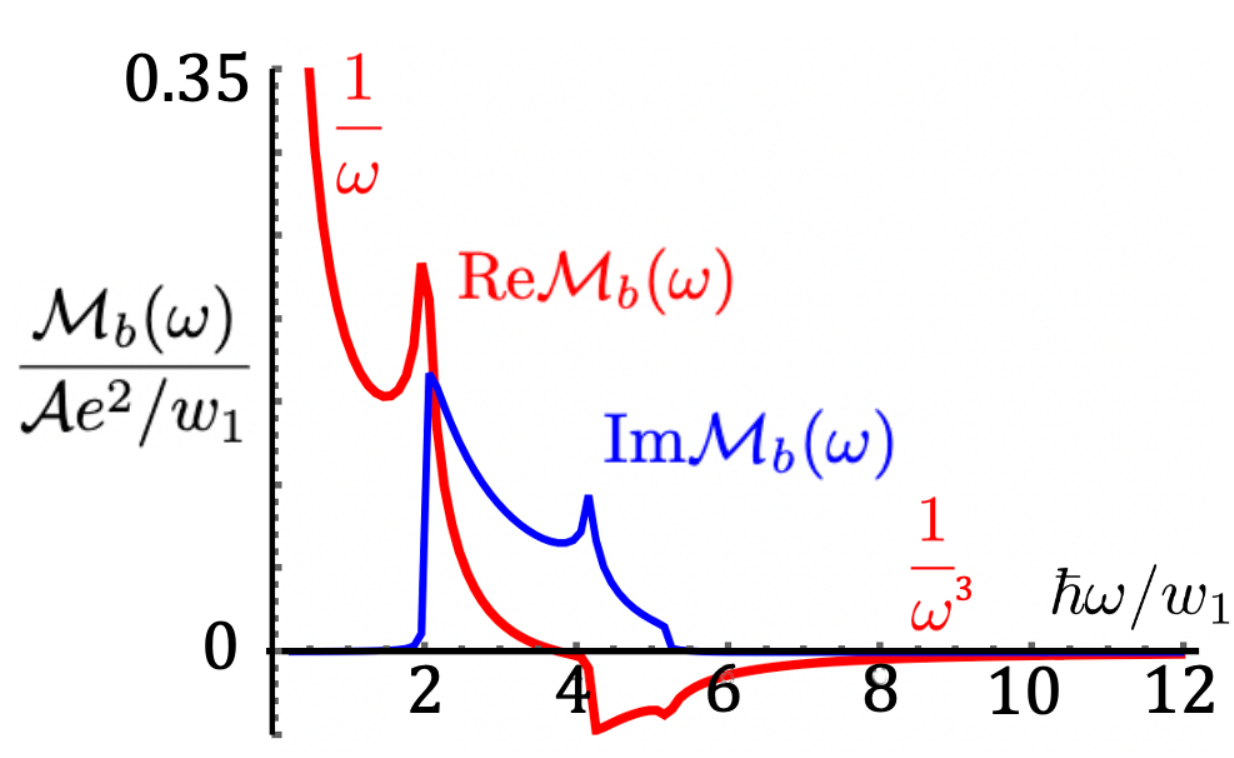} \\
	\vspace{0.5cm}
	(b)\includegraphics[width=0.9\linewidth]{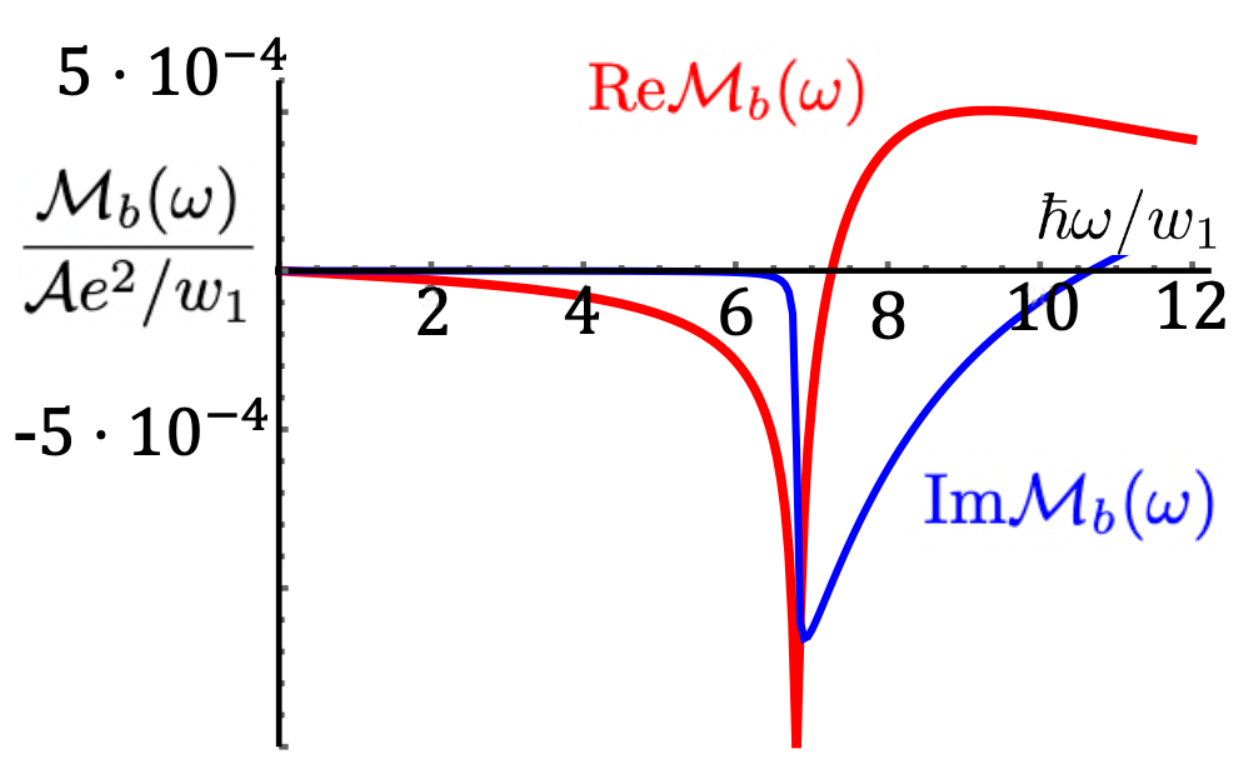} 	
\caption{The real and imaginary parts of $\mathcal M_b(\omega)$ from Eq.~(\ref{M_Berry_app}) vs frequency $\omega$ for (a) topological ($C=1$) and (b) trivial ($C=0$) phases of the Haldane model.   The vertical scale indicates that $\mathcal M_b(\omega)$ is much weaker in the trivial phase.} 
	\label{fig:M_vs_omega}
\end{figure}
%%%%%%%%%%%%%%%%%%%%%%%%%%%%%%%%%%%%%%%%%%%%%%%%%%%%%%%%%%%%%%%%%%%%%%%%%%%%
Let us illustrate the results of our paper for a model of the Chern insulator proposed by Haldane \cite{Haldane1988}. The model is formulated on the honeycomb lattice shown in Fig.~\ref{fig:lattice_and_spectrum}(a). The lattice sites are connected by the tunneling amplitudes $w_1$ along the nearest-neighbor vectors $\bm a_1,\bm a_2$ and $\bm a_3$, as well as by the purely imaginary tunneling amplitude $iw_2$ along the next-nearest-neighbor vectors $\bm b_1, \bm b_2$ and $\bm b_3$. We also introduce a staggered on-site potential of opposite signs $\pm\Delta$ on the two sublattices.  In momentum space, the corresponding Hamiltonian is a $2\times2$ matrix, because the honeycomb lattice is bipartite. It is practical to expand the Hamiltonian in Pauli matrices $\bm \sigma =(\sigma_x,\sigma_y,\sigma_z)$
\begin{align}
  & H(\bm k) = \bm \sigma\cdot \bm w(\bm k), \qquad \bm w(\bm k)= \sum_{j=1,2,3} 
  \label{hamiltonian_haldane} \\
  &\left[w_1\cos(\bm k\cdot\bm a_j), \,\,
  w_1 \sin(\bm k\cdot\bm a_j), \,\, \Delta+ 2w_2 \sin(\bm k\cdot \bm b_j)\right], 
  \nonumber
\end{align}
parametrized by the momentum-dependent vector $\bm w(\bm k)$. The energy spectrum of the Hamiltonian consists of the two bands $\varepsilon_{c,v}(\bm k) = \pm |\bm w(\bm k)|$. For the specific choice of hopping parameters $w_2 =  w_1/2$ and $\Delta = 0$, the bands span the energy range $w_1<|\varepsilon_{c,v}(\bm k)|<3w_1$ and are separated by the gap $\Delta=2w_1$.
The Haldane model has a nonvanishing Berry curvature~\cite{Yakovenko1990} $\Omega_{v}(\bm k) = -\Omega_{c}(\bm k) = {\bm w(\bm k)\cdot [\partial^x\bm w(\bm k) \times \partial^y \bm w(\bm k)]/2|\bm w(\bm k)|^3}$ shown by color in Fig.~\ref{fig:lattice_and_spectrum}(b). The dc Hall conductivity is quantized $\sigma_H = Ce^2/h$, where the integer Chern number $C$ is the topological invariant given by Eq.~(\ref{Chern_app}).  The competition between the hopping amplitude $w_2$ and the on-site potential $\Delta$ determines the topological regime of the model. The system is topologically non-trivial with $|C|=1$ if $3\sqrt 3|w_2|>|\Delta|$ and is topologically trivial with $C=0$ otherwise.

We numerically evaluate $\mathcal M_b(\omega)$ versus frequency $\omega$ for the Haldane model~(\ref{hamiltonian_haldane}) using Eq.~(\ref{M_Berry_app}).   The results are presented in Fig.~\ref{fig:M_vs_omega}.  The two panels differ in the magnitude of the staggered on-site potential: (a) $\Delta = 0$ and (b) $\Delta = 6w_1$, and correspond to the topological $C=1$ and trivial $C=0$ phases.  Both real and imaginary parts of $\mathcal M_b(\omega)$ are shown, related by the appropriate Kramers-Kronig relation.  The former describes the asymmetric ac Stark effect energy shift, whereas the latter represents photon absorption due to resonant interband transitions.  Interband transitions occur in the frequency range $2w_1<\hbar\omega<6w_1$ at the locus of points in the momentum space where the resonant condition $|\varepsilon_2(\bm k)-\varepsilon_1(\bm k)| = \hbar\omega$ is satisfied.   Both the real and imaginary parts of $\mathcal M_b(\omega)$ exhibit a series of kinks and dips in the resonant region associated with band edges and van Hove singularities. For this reason, Re$\mathcal M_b(\omega)$ has logarithmic divergences induced by the band edges at $\hbar\omega/w_1 = 2$ and $\hbar\omega/w_1 \approx 5$ in Fig.~\ref{fig:M_vs_omega}(a). The $1/\omega$ divergence at low $\omega$ is canceled by the edge contribution not included here, as discussed in Sec.~\ref{sec:edge}.   In contrast, Fig.~\ref{fig:M_vs_omega}(b) shows that $\mathcal M_b(\omega)$ is much weaker in the topologically trivial phase, but is still nonzero.

%%%%%%%%%%%%%%%%%%%%%%%%%%%%%%%%%%%%%%%%%%%%%%%%%%%%%%%%%%%%%%%%%%%%%%%%%%%%%
\bibliography{biblio}
%%%%%%%%%%%%%%%%%%%%%%%%%%%%%%%%%%%%%%%%%%%%%%%%%%%%%%%%%%%%%%%%%%%%%%%%%%%%%

%%%%%%%%%%%%%%%%%%%%%%%%%%%%%%%%%%%%%%%%%%%%%
%%%%%%%%%%%%%%%%%%%%%%%%%%%%%%%%%%%%%%%%%%%%%
\end{document}